\documentclass[letterpaper,11pt]{article}

\usepackage{jheppub}
\usepackage{graphicx}
\usepackage{amsmath}
\usepackage{amssymb,empheq}
\usepackage{mathrsfs}
\usepackage{bm}
\usepackage{braket}
\usepackage{slashed}
\usepackage{xspace}
\usepackage[dvipsnames]{xcolor}
\usepackage{mdwlist}
\usepackage[normalem]{ulem}
\usepackage[utf8]{inputenc}

\newcommand{\eq}[1]{eq.~\eqref{eq:#1}}
\newcommand{\eqs}[2]{eqs.~\eqref{eq:#1} and \eqref{eq:#2}}
\renewcommand{\sec}[1]{sec.~\ref{sec:#1}}

\newcommand{\fig}[1]{fig.~\ref{fig:#1}}

\newcommand{\nn}{\nonumber \\ }

\def\bea{\begin{eqnarray}}
\def\eea{\end{eqnarray}}
\def\xbj{x_{\rm bj}}

\newcommand{\df}{\mathrm{d}}
\newcommand{\rd}{\mathrm{d}}
\newcommand\mathd{\mathrm{d}} 
\newcommand{\img}{\mathrm{i}}
\newcommand{\sdt}{\!\cdot\!}

\newcommand{\lra}{\leftrightarrow}

\newcommand{\al}{\alpha}

\newcommand{\cL}{{\mathcal L}}

\newcommand{\bn}{{\bar{n}}}

\newcommand\smu{\left(\mathcal{S}\mu\right)}
\def\W{\mathcal{W}}

\newcommand{\MSbar}{\ensuremath{\overline{\text{MS}}}}
\def\qed{U(1)_{\text{em}}}
\newcommand\sigmaprobe[1]{{\sigma_{{l#1}}}}

\newcommand\xph{x}

\allowdisplaybreaks[2]

\preprint{\vbox{\hbox{Nikhef 2018-011}}}

\title{Electroweak Gauge Boson Parton Distribution Functions}

\author[a]{Bartosz Fornal,}
\author[a]{Aneesh V.~Manohar,}
\author[b,c]{Wouter J.~Waalewijn}

\affiliation[a]{Department of Physics, University of California, San Diego, 9500 Gilman Drive, La Jolla, CA 92093, USA}
\affiliation[b]{Institute for Theoretical Physics Amsterdam and Delta Institute for Theoretical Physics, University of Amsterdam, Science Park 904, 1098 XH Amsterdam, The Netherlands}
\affiliation[c]{Nikhef, Theory Group, Science Park 105, 1098 XG, Amsterdam, The Netherlands}

\abstract{Transverse and longitudinal electroweak gauge boson parton distribution functions (PDFs) are computed in terms of deep-inelastic scattering structure functions, following the recently developed method to determine the photon PDF. The calculation provides  initial conditions at the electroweak scale for PDF evolution to higher energies. Numerical results for the $W^\pm$ and $Z$ transverse, longitudinal and polarized PDFs, as well as the $\gamma Z$ transverse and polarized PDFs are presented.}
\begin{document}
\maketitle

%%%%%%%%%%%%%%%%%%%%%%%%%%%%%%%%%%%%%%%%%%%%%%%%%%%%%%%%%%%%%%%%%%%%%%%%%%%%%%%%
\section{Introduction}
\label{sec:intro}
%%%%%%%%%%%%%%%%%%%%%%%%%%%%%%%%%%%%%%%%%%%%%%%%%%%%%%%%%%%%%%%%%%%%%%%%%%%%%%%%

An essential ingredient in calculations of high energy scattering cross sections are the parton distribution functions (PDFs), which describe the incoming protons. These usually only encode QCD effects, but at the multi-TeV energies probed by collisions at the Large Hadron Collider, electroweak effects start becoming important. At Future Circular Collider energies, electroweak effects are order one~\cite{Mangano:2016jyj}, because of  Sudakov double logarithms in the electroweak PDF evolution~\cite{Ciafaloni:2000df,Ciafaloni:2000rp}, which are absent for QCD. This difference is due to the spontaneous breaking of electroweak symmetry, implying that PDFs only have to be QCD (and QED) singlets, but not necessarily electroweak singlets. Indeed, it is the $SU(2) \times U(1)$ non-singlet PDFs that have Sudakov double logarithms in their evolution. 

Electroweak contributions to PDF evolution have been computed recently~\cite{Bauer:2017bnh,Bauer:2017isx,Manohar:2018kfx}, which relates PDFs at different scales. However, the PDFs themselves have to be determined from experiment. Recently, the photon PDF was calculated directly in terms of deep-inelastic scattering structure functions~\cite{Manohar:2016nzj,Manohar:2017eqh}. In this paper, we use a similar method to compute the $W$ and $Z$ PDFs. Massive gauge bosons have both transverse and longitudinal polarizations, and a new feature of our analysis is the computation of PDFs for longitudinally polarized gauge bosons. In contrast to the photon PDF, nonperturbative contributions are suppressed, allowing us to calculate the gauge boson PDFs in terms of quark PDFs at the electroweak scale.

Section~\ref{sec:transverse} computes the transverse and polarized $W^\pm$, $Z$ and $\gamma Z$ gauge boson PDFs (which are the sum and difference of the helicity $h=\pm 1$ PDFs) using operator methods. The $W^\pm$ and $Z$ longitudinal PDFs, i.e.~$h=0$, are computed in \sec{longitudinal}. In \sec{compare} we compare our results with previous ones in the literature based on the effective $W$ approximation~\cite{Chanowitz:1984ne,Dawson:1984gx,Kane:1984bb}. We present an alternative derivation in sec.~\ref{sec:factorization} using factorization methods.
Numerical values for the PDFs are presented in sec.~\ref{sec:numerics}.

%%%%%%%%%%%%%%%%%%%%%%%%%%%%%%%%%%%%%%%%%%%%%%%%%%%%%%%%%%%%%%%%%%%%%%%%%%%%%%%%
\section{Transverse gauge boson PDFs}
\label{sec:transverse}
%%%%%%%%%%%%%%%%%%%%%%%%%%%%%%%%%%%%%%%%%%%%%%%%%%%%%%%%%%%%%%%%%%%%%%%%%%%%%%%%

We start this section with defining the PDFs of transverse gauge bosons. We then derive how these are related to structure functions in deep-inelastic scattering. Evaluating the structure functions to lowest order in the strong coupling $\al_s$, we obtain a formula in terms of the quark PDFs.

%===============================================================================
\subsection{Definition}
%===============================================================================

We start by briefly reviewing the PDF definition for quarks and gluons in QCD, before discussing the electroweak gauge boson case. We will frequently use light-cone coordinates,  decomposing a four-vector $p^\mu$ as
%%%
\begin{align}
  p^\mu = p^-\, \frac{n^\mu}{2} + p^+\, \frac{\bn^\mu}{2} + p_\perp^\mu
  \,, \qquad
  p^- = \bn \sdt p
  \,, \qquad
  p^+ = n \sdt p 
\,,\end{align}
%%%
where $n^\mu=(1,0,0,1)$ and $\bar n^\mu =(1,0,0,-1)$ are two null vectors with $n \cdot \bar n=2$ and $p_\perp$ is transverse to both $n^\mu$ and $\bn^\mu$. The QCD PDF operators are defined as~\cite{Collins:1981uw}
%%%
\begin{align}\label{eq:cs}
O_Q(r^-) &= \frac{1 }{ 4 \pi} \int_{-\infty}^\infty \! \df \xi\, e^{-\img \xi r^-}
[ \bar Q(\bar n \xi )\, \W (\bar n \xi )]\ \slashed{\bar n}\ [\W^\dagger (0) \,  Q(0 ) ] \,,\nn
O_G(r^-) &= -\frac{1 }{ 2 \pi r^-} \int_{-\infty}^\infty\! \df  \xi\, e^{-\img \xi r^-}
 \bn_\mu [G^{\mu \lambda }(\bar n \xi)\, \W (\bar n \xi )]\ \bn_\nu [\W^\dagger (0) \,G^{\nu}{}_\lambda(0)]\,,
\end{align}
%%%
for quarks and gluons, respectively. Here $\W$ is a Wilson line,\footnote{It is conventional to use $\W^\dagger (x) \,  Q(x)$ for the field, so the Wilson line must end at $x$. We use the sign convention $D_\mu = \partial_\mu + i g A_\mu$. }
\begin{align}
 \W(x) = {P} \exp\bigg\{-\img\, g \int_{\infty}^0\!\df s\, \bn \sdt \big[ A(x+s \bar n)\big]\bigg\}\,,
 \end{align}
along the $\bar n$ direction in the fundamental representation for the quark, and in the adjoint representation for the gluon PDF operator, ensuring gauge invariance. For the anti-quark,  $Q \lra \bar Q$ and the Wilson line is in the anti-fundamental representation. The PDF operators involve an ordinary product of fields, not the time-ordered product, so the Feynman rules are those for cut graphs. The quark and gluon PDFs are given by the matrix elements of these operators in a proton state of momentum $p$, 
%%%
\begin{align}\label{eq:pdf}
f_{Q}(r^-/p^-,\mu) &\equiv \braket{p| O_Q(r^-) |p}_c
\,,&
f_{G}(r^-/p^-,\mu) &\equiv  \braket{p| O_G(r^-) |p}_c
\,,\end{align}
%%%
where only connected graphs contribute. 

In the Standard Model (SM) at high energies, fermion PDFs are defined in terms of the $SU(2) \times U(1)$ fields $q$, $\ell$, $u$, $d$, $e$, 
where $q$, $\ell$ are left-handed $SU(2)$ doublet fields, and $u$, $d$, $e$ are right-handed $SU(2)$ singlet fields. The QCD Wilson line is replaced by a $SU(3) \times SU(2) \times U(1)$ Wilson line in the representation of the fermion field. The new feature in the electroweak case is that the PDF operator does not have to be a $SU(2) \times U(1)$ singlet. In particular, for the quark doublet $q$ there are two operators,
%%%
\begin{align} \label{eq:quark}
O^{(1)}_q(r^-) &= \frac{1 }{ 4 \pi} \int_{-\infty}^\infty \df \xi\, e^{-\img \xi r^-}\,
[ \bar q(\bar n \xi ) \, \W (\bar n \xi )]_{i\alpha} \ \slashed{\bar n}\
\, \delta^i_j\,  [\W^\dagger (0) \,  q(0 ) ]^{j \alpha}
\,,  \nn
O^{(\text{adj},a)}_q(r^-) &= \frac{1 }{ 4 \pi}\int_{-\infty}^\infty \df \xi\, e^{-\img \xi r^-}\, [ \bar q(\bar n \xi )\, \W (\bar n \xi )]_{i\alpha} \ \slashed{\bar n}\
 [t^a]^i{}_j\,  [\W^\dagger (0) \,  q(0 ) ]^{j \alpha}
\,, \end{align}
%%%
where $i,j$ are gauge indices in the fundamental representation of $SU(2)$, $t^a$ is an $SU(2)$ generator, and $\alpha$ is a gauge index in the fundamental representation of $SU(3)$. $O_q^{(1)}$ is an $SU(3) \times SU(2) \times U(1)$ singlet, but $O_q^{(\text{adj},a)}$ is an $SU(3) \times U(1)$ singlet, and transforms as an $SU(2)$ adjoint. The proton matrix elements of the operators give the $u_L$ and $d_L$ PDFs,
%%%
\begin{align}\label{eq:ud}
f_{u_L}(r^-/p^-,\mu)&= \braket{p|\, \tfrac12 O^{(1)}_q(r^-) +  O^{(\text{adj},a=3)}_q(r^-)\, |p}
\,, \nn
f_{d_L}(r^-/p^-,\mu)&= \braket{p|\, \tfrac12 O^{(1)}_q(r^-) -  O^{(\text{adj},a=3)}_q(r^-)\, |p}
\,.\end{align}
%%%
Since electroweak symmetry is broken, $O^{(\text{adj},a=3)}_q$ can have a non-zero matrix element in the proton, such that $f_{u_L} \neq f_{d_L}$. The evolution above the electroweak scale of $O^{(1)}_q$, $O^{(\text{adj},a)}_q$ and of the corresponding gauge and Higgs boson operators was computed in ref.~\cite{Manohar:2018kfx}. 

The quark PDF operators in \eq{quark} in the unbroken theory can be matched onto PDF operators in the broken theory at the electroweak scale. At tree level this matching is trivial,
%%%
\begin{align} \label{eq:low_matching}
O^{(1)}_q(r^-) &= O_{u_L}(r^-) + O_{d_L}(r^-)
\,, \nn
O^{(\text{adj},a=3)}_q(r^-)  &= \tfrac12 O_{u_L}(r^-) -\tfrac12 O_{d_L}(r^-) 
\,,\end{align}
%%%
where
%%%
\begin{align} \label{eq:uL}
O_{u_L}(r^-) &= \frac{1 }{ 4 \pi} \int \df \xi\, e^{-\img \xi r^-}\,
[ \bar u_L(\bar n \xi ) \, \W (\bar n \xi )]_{\alpha} \ \slashed{\bar n}\
\,   [\W^\dagger (0) \,  u_L(0 ) ]^{\alpha}
\,, \end{align}
%%%
and similarly for $d_L$.  Essentially all we have done is replace the $SU(3) \times SU(2) \times U(1)$ Wilson lines by $SU(3) \times \qed$ Wilson lines, so $\W$ in \eq{uL} only contains gluons and photons.

The gauge PDF operators in irreducible $SU(2)$ representations were given in ref.~\cite{Manohar:2018kfx}. At lowest order in electroweak corrections, the matching onto the broken operators is analogous to \eq{low_matching} and given in eq.~(5.1) of  ref.~\cite{Manohar:2018kfx}. The relevant PDF operators in the broken theory are\footnote{\label{foot1} Note that $W$ is the $SU(2)$ gauge field, and $\W$ is the Wilson line. We have switched conventions relative to ref.~\cite{Manohar:2017eqh}, $n \leftrightarrow \bar n$, $p^+ \leftrightarrow p^-$. The photon PDF operator in ref.~\cite{Manohar:2017eqh} was written as the sum of two terms, such that it has manifest antisymmetry under $x \to -x$. However, the commutator of light-cone operators does not contribute to the connected matrix element~\cite{Jaffe:1983hp}, so the two terms can be combined into a single term shown in \eq{W}. The two terms in $O_{\Delta \gamma}$, etc.\ can be similarly combined.}
%%%
\begin{align}\label{eq:W}
O_{W^+_T}(r^-) &= -\frac{1 }{ 2 \pi r^-} \int_{-\infty}^\infty \! \df  \xi\, e^{-\img \xi r^-}\,
\bn_\mu [W^{-\mu \lambda }(\bar n \xi)\, \W (\bar n \xi )]\,   \bn_\nu [\W^\dagger (0) \,W^{+\nu}{}_\lambda(0)]\,, \nn
O_{W^-_T}(r^-) &= -\frac{1 }{ 2 \pi r^-} \int_{-\infty}^\infty \! \df  \xi\, e^{-\img \xi r^-}\,
\bn_\mu [W^{+\mu \lambda }(\bar n \xi)\, \W (\bar n \xi )]\,   \bn_\nu [\W^\dagger (0) \,W^{-\nu}{}_\lambda(0)]\,, \nn
O_{\gamma}(r^-) &= -\frac{1 }{ 2 \pi r^-} \int_{-\infty}^\infty \! \df  \xi\, e^{-\img \xi r^-}\,
\bn_\mu F^{\mu \lambda }(\bar n \xi)\,   \bn_\nu F^{\nu}{}_\lambda(0)\,, \nn
O_{Z_T}(r^-) &= -\frac{1 }{ 2 \pi r^-} \int_{-\infty}^\infty\! \df  \xi\, e^{-\img \xi r^-}\,
\bn_\mu Z^{\mu \lambda }(\bar n \xi)\,   \bn_\nu Z^{\nu}{}_\lambda(0)\,, \nn
O_{Z_T\gamma}(r^-) &= -\frac{1 }{ 2 \pi r^-}\int_{-\infty}^\infty \! \df  \xi\, e^{-\img \xi r^-}\,
\bn_\mu Z^{\mu \lambda }(\bar n \xi)\,   \bn_\nu  F^{\nu}{}_\lambda(0)\,, \nn
O_{\gamma Z_T}(r^-) &= -\frac{1 }{ 2 \pi r^-}\int_{-\infty}^\infty\! \df  \xi\, e^{-\img \xi r^-}\,
\bn_\mu F^{\mu \lambda }(\bar n \xi)\,   \bn_\nu  Z^{\nu}{}_\lambda(0)\,, 
\end{align}
%%%
in terms of the field-strength tensors. Note that the PDFs $f_{Z_T\gamma}$ and $f_{\gamma Z_T}$ are related by complex conjugation. The PDF operators in \eq{W} are invariant under $SU(3) \times \qed$ gauge transformations. For $O_{W_T^+}$ this involves a $\qed$ Wilson line $\W$ with $Q=1$, and  $O_{W_T^-}$ has $Q=-1$.
There are no Wilson lines for $\gamma$ and $Z$, since they are neutral.

As we now show, the operators in \eq{W} only capture the transverse polarizations.
A gauge boson moving in the $n$ direction has momentum and polarization vectors
%%%
\begin{align}\label{eq:2.8a}
k^\mu &= (E_k,0,0,k), & \epsilon_+^\mu &= -\frac{1}{\sqrt 2} (0,1,\img,0),  & \epsilon_{-}^\mu &= \frac{1}{\sqrt 2} (0,1,-\img,0),
& \epsilon_0^\mu &= \frac{1}{M} (k,0,0,E_k),
\end{align}
%%%
which satisfy $k \cdot \epsilon_\lambda =0$ and $\epsilon_\lambda^* \cdot \epsilon_\sigma = - \delta_{\lambda \sigma}$.
By calculating the matrix element of the field-strength tensors appearing in \eq{W} for a gauge boson state,
%%%
\begin{align}\label{eq:2.8}
\braket{k,\epsilon | \bn_\mu F^{\mu \lambda }(\bar n \xi)\,   \bn_\nu F^{\nu}{}_\lambda(0)| k, \epsilon}
&=\left[ (\bar n \cdot k)^2 (\epsilon^* \cdot \epsilon) + (\bar n \cdot \epsilon) (\bar n \cdot \epsilon^*)k^2 \right] e^{\img (\bar n \cdot k) \xi} \nn
&=  -(\bar n \cdot k)^2 e^{\img (\bar n \cdot k) \xi} \times \begin{cases} 1, & \epsilon=\epsilon_+, \epsilon_-\,, \\
0, &\epsilon=\epsilon_0\,,
\end{cases}
\end{align}
%%%
we conclude that the PDF operators only pick out transversely-polarized gauge bosons. 

The transverse PDFs $f_{W^+_T}$, etc.~defined through the operators in \eq{W}, sum over the helicity $h=\pm 1$ contributions. The longitudinal gauge boson PDF encodes the $h=0$ contribution, and will be discussed in \sec{longitudinal}. In addition, we will also consider the polarized $W^+$ PDF, 
\begin{align}
f_{\Delta W_T^+}=f_{W^+(h=1)}-f_{W^+(h=-1)}\,, 
\end{align}
etc.  In an unpolarized proton target, the gluon distribution $f_{\Delta g}$ vanishes, as can be shown by reflecting in the plane of the incident proton. However, the weak interactions violate parity, so $f_{\Delta W^\pm_T}$ and $f_{\Delta Z_T}$ do not vanish. The polarized photon PDF can be written as~\cite{Manohar:1990kr,Manohar:1990jx} (see footnote \ref{foot1}), 
%%%
\begin{align}\label{eq:2.9}
O_{\Delta \gamma}(r^-) &= \frac{\img}{ 2 \pi r^-} \int_{-\infty}^\infty \! \df  \xi\, e^{-\img \xi r^-}\,
\bn_\mu F^{\mu \lambda }(\bar n \xi)\,   \bn_\nu \widetilde F^{\nu}{}_\lambda(0)\,, 
\end{align}  
%%%
where $\widetilde F_{\alpha \beta} = \frac12 \epsilon_{\alpha \beta \lambda \sigma} F^{\lambda \sigma}$ with $\epsilon_{0123}=+1$, and we use the 't~Hooft-Veltman convention for the $\epsilon$-symbol and $\gamma_5$. Similar expressions hold for the polarized versions of the other PDFs in \eq{W}.

%===============================================================================
\subsection{Evaluation}
\label{sec:eva_T}
%===============================================================================

\begin{figure}
\begin{align*}
{\setlength{\arraycolsep}{0.15cm}
\begin{array}{ccc}
\begin{minipage}{3.5cm} \includegraphics[width=3.5cm]{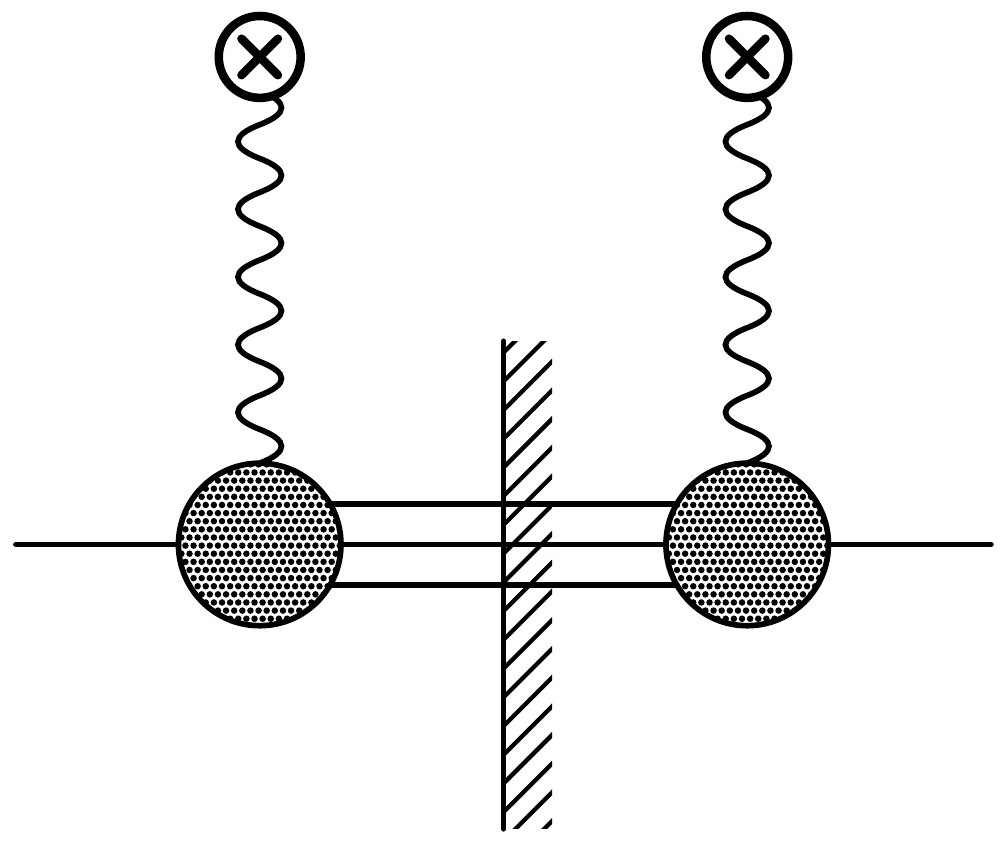} \end{minipage} &  &
\begin{minipage}{3.5cm} \includegraphics[width=3.5cm]{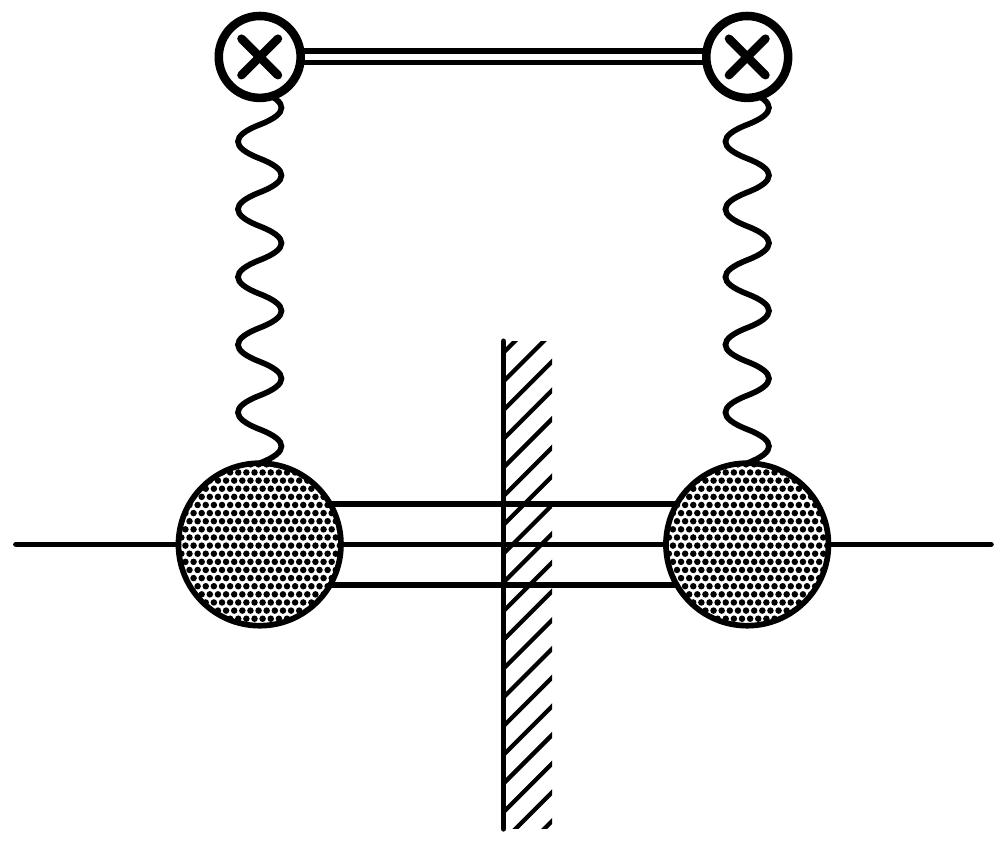} \end{minipage} \\[5pt]
(a) && (b)
\end{array}
}
\end{align*}
\caption{\label{fig:pdfop} Matrix element of the PDF operator in a proton state for (a) photon and $Z$ PDFs,  and (b)~$W$~PDFs,  where the $\qed$ Wilson line is shown as a double line. The $\otimes$ vertex is the field-strength tensor and the bottom part of the graphs is the hadronic tensor $W_{\mu \nu}(p,q)$.}
\end{figure}

We now discuss how the transverse gauge boson PDFs can be computed from \fig{pdfop}, following the procedure in refs.~\cite{Manohar:2016nzj,Manohar:2017eqh}. 
We start by introducing the hadronic tensor and structure functions, briefly repeat the argument for the photon case, and then generalize to the other gauge boson PDFs in \eq{W}. Only PDFs for unpolarized proton targets will be considered, but it is straightforward to generalize to polarized protons.
%%% FIG %%%
%%%

\begin{figure}
\centering
\includegraphics[width=3.5cm]{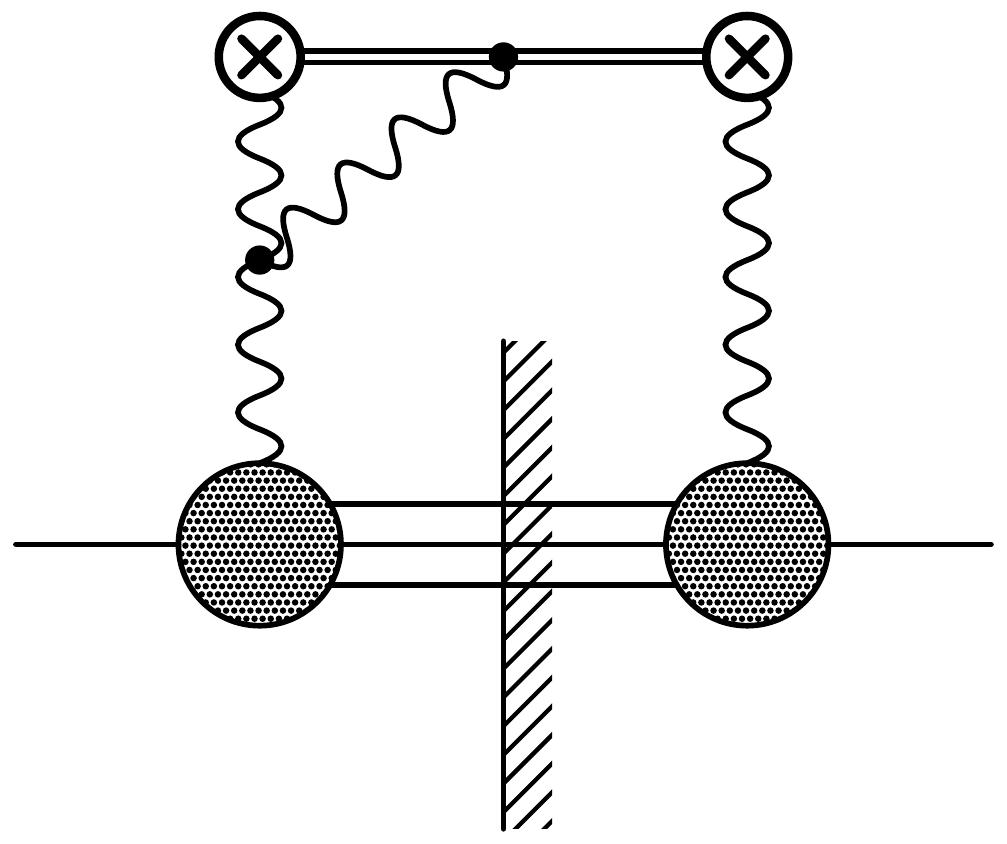}\hspace{1cm}
\includegraphics[width=3.5cm]{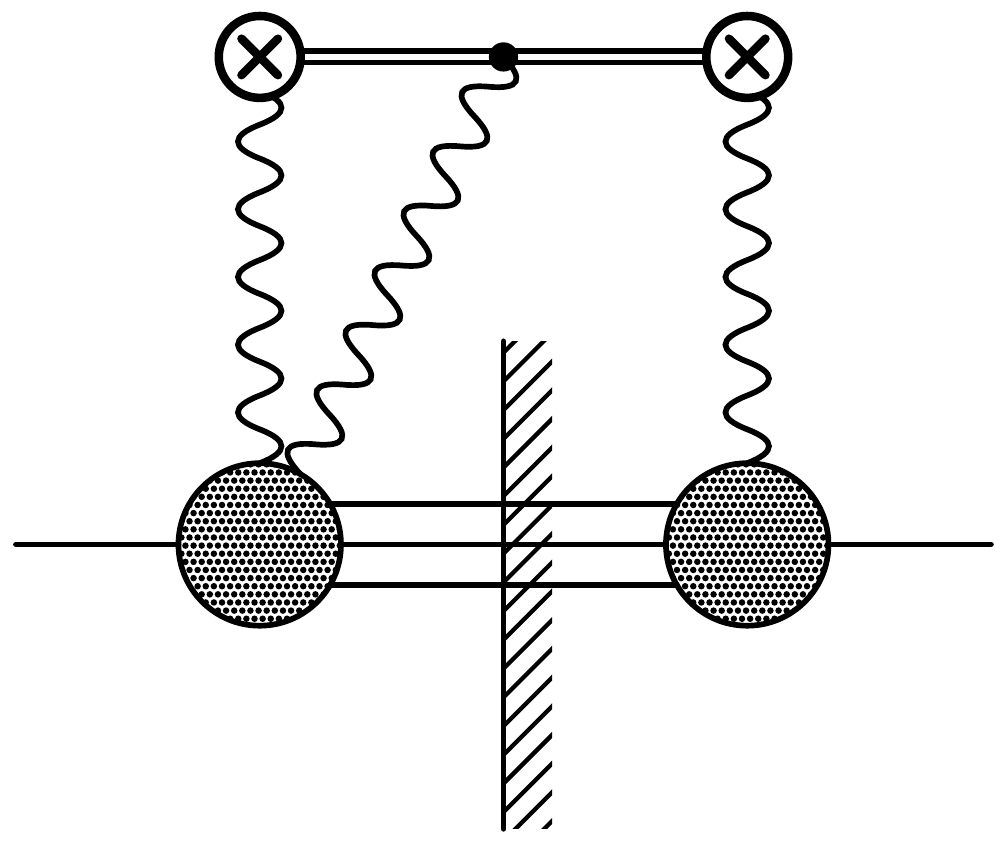}
\caption{\label{fig:2} Radiative corrections to $W$ boson PDFs. The right diagram does not allow a simple factorization in terms of structure functions, as it involves a three-point correlator in the proton.}
\end{figure}

The electroweak PDFs at high energies evolve using anomalous dimensions in the unbroken theory computed in refs.~\cite{Bauer:2017isx,Manohar:2018kfx}, which contain Sudakov double logarithms. In this paper, we compute the initial conditions to this evolution at the electroweak scale. Since the electroweak gauge bosons are massive, the logarithmic evolution is not important until energies well above the electroweak scale. 
In addition, there are radiative corrections for the $W$ PDFs from interactions with the Wilson line from graphs shown in \fig{2}, which are absent in the photon case.
For this reason, we compute the electroweak PDFs to order $\alpha \sim \alpha_2 \sim \alpha_Z$. 
%%% FIG %%%
%%%

The lower part of the graph in \fig{pdfop} is the hadronic tensor defined as
%%%
\begin{align}\label{eq:wmunu}
W_{\mu\nu}(p,q) = \frac{1}{4\pi}\int \rd^4 z \ e^{\img q \cdot z}
\langle p | \bigl[ j^\dagger_\mu(z), j_\nu(0)\bigr] | p \rangle \,,
\end{align}
%%%
where $p$ is the proton momentum and $q$ is the incoming gauge-boson momentum at vertex $j_\nu$. The standard decomposition of $W_{\mu \nu}$ is
%%%
\begin{align}\label{eq:sf}
W_{\mu\nu}(p,q) &= F_1 \Big(-g_{\mu\nu} + {q_\mu
q_\nu\over q^2}\Big) + {F_2\over p \cdot q} \Big(p_\mu - {p\cdot q \ q_\mu\over
q^2}\Big)
\Big(p_\nu - {p\cdot q\ q_\nu\over q^2}\Big) - {\img F_3 \over 2 p\cdot q}\ \epsilon_{\mu\nu\lambda\sigma}
q^\lambda p^\sigma ,
\end{align}
%%%
in terms of the structure functions $F_1$, $F_2$, $F_3$, which depend on the Bjorken variable 
%%%
\begin{align}\label{eq:3.22}
\xbj &= \frac{Q^2}{2 p \cdot q}
\,,\end{align}
%%%
and $Q^2=-q^2$. It is convenient to replace $F_1$ in our results by the longitudinal structure function,
%%%
\begin{align} \label{eq:3.28}
  F_L(\xbj,Q^2) &\equiv \bigg(1+\frac{4 \xbj^2m_p^2}{Q^2}\bigg)F_2(\xbj,Q^2)  - 2\xbj F_1(\xbj,Q^2) 
\,.\end{align}
%%%

The currents in \eq{wmunu} depend on the process. For the photon PDF, $j_\mu$ is the electromagnetic current and $F_3$ vanishes. For $W$ PDFs, $j_\mu$ is the weak charged current, and for $Z$ PDFs, $j_\mu$ is the weak neutral current. These currents follow from the interaction Lagrangian, which is given by~\cite{Patrignani:2016xqp} 
%%%
\begin{align}\label{eq:3.29}
\mathcal{L}_{\text{int}} &= -  e A_\mu j^\mu_{\text{em}} -  \frac{g_2}{2 \sqrt 2} \big( W^+_\mu j^\mu_W+W^-_\mu j^{\dagger \mu}_W  \big) -  \frac{g_Z}{2} Z_\mu j^\mu_Z ,
\end{align}
%%%
using the conventional normalization of currents in deep-inelastic scattering.
Here $g_2=e/\!\sin\theta_W$ and $g_Z=e/(\sin\theta_W \cos \theta_W)$ and
%%%
\begin{align}
j^\mu_{\text{em}} &= \tfrac23 \bar u \gamma^\mu u + \ldots, \nn
j^\mu_W &= V_{ud}\, \bar u \gamma^\mu(1-\gamma_5) d + \ldots, \nn
j^\mu_Z &= \bar u \left[ \gamma^\mu \left(\tfrac12 - \tfrac43 \sin^2\theta_W \right) - \tfrac12\gamma^\mu \gamma_5 \right] u + \ldots \,.
\end{align}
%%%

The structure functions for electromagnetic scattering  are denoted by $F_i^{(\gamma)}$, for neutrino scattering $\nu p \to e^- X$ by $F_i^{(\nu)}$, for anti-neutrino scattering  $\bar \nu p \to e^+ X$ by $F_i^{(\bar \nu)}$, for neutral current scattering by $F_i^{(Z)}$, and for $\gamma-Z$ interference by $F_i^{(\gamma Z)}$ and $F_i^{(Z \gamma)}$, where the superscripts indicate the $j_\mu$ and $j_\nu$ current used in \eq{wmunu}. In QCD, to lowest order in $\alpha_s$, $F_L=0$  and
%%%
\begin{align}\label{eq:2.17}
{F}^{(\nu)}_2(x,Q^2) &= 4 x [ f_{d_L}(x,Q) + f_{\bar u_R}(x,Q) ], &
{F}^{(\nu)}_3(x,Q^2) &= 4  [ f_{d_L}(x,Q) - f_{\bar u_R}(x,Q) ],  \nn
{F}^{( \bar \nu )}_2(x,Q^2) &= 4 x [ f_{u_L}(x,Q) +  f_{\bar d_R}(x,Q) ] ,&
{F}^{( \bar \nu)}_3(x,Q^2) &= 4 [ f_{u_L}(x,Q) - f_{\bar d_R}(x,Q) ] , 
\end{align}
%%%
and 
%%%
\begin{align}\label{eq:2.17c}
F^{(Z)}_2(x,Q^2) &=  4x \sum_{q=u,d} g_{Lq}^2 [f_{q_L}(x,Q)+f_{\bar q_R}(x,Q)] + g_{Rq}^2 [ f_{q_R}(x,Q) + f_{\bar q_L}(x,Q) ]  , \nn
F^{(Z)}_3(x,Q^2) &= 4 \sum_{q=u,d} g_{Lq}^2 [f_{q_L}(x,Q)-f_{\bar q_R}(x,Q)] - g_{Rq}^2 [ f_{q_R}(x,Q) - f_{\bar q_L}(x,Q) ], \nn
F^{(\gamma Z)}_2(x,Q^2) &=F^{(Z\gamma )}_2(x,Q^2) \nn 
&=  2x \sum_{q=u,d} g_{Lq}\, \mathcal{Q}_q [f_{q_L}(x,Q)+f_{\bar q_R}(x,Q)] + g_{Rq} \, \mathcal{Q}_q [ f_{q_R}(x,Q) + f_{\bar q_L}(x,Q) ] , \nn
F^{(\gamma Z)}_3(x,Q^2) &=F^{(Z\gamma )}_3(x,Q^2) \nn
&= 2\sum_{q=u,d} g_{Lq}\, \mathcal{Q}_q [f_{q_L}(x,Q)-f_{\bar q_R}(x,Q)] - g_{Rq} \, \mathcal{Q}_q [ f_{q_R}(x,Q) - f_{\bar q_L}(x,Q) ]  
\,.\end{align}
%%%
Here the subscripts $L,R$ denote the parton helicities, not chiralities, $\mathcal{Q}_q$ is the electric charge, 
%%%
\begin{align}
g_{Lu}&=\tfrac12 - \tfrac23 \sin^2 \theta_W\,, &
g_{Ru}&=- \tfrac23 \sin^2 \theta_W\,, \nn
g_{Ld}&=-\tfrac12 + \tfrac13 \sin^2 \theta_W\,, &
g_{Rd}&= \tfrac13 \sin^2 \theta_W
\,,\end{align}
%%%
and we have neglected CKM mixing and heavier quark flavors.
For an unpolarized proton beam, the expressions can be simplified using $f_{u_L}=f_{u_R}=\tfrac12 f_u$, etc.

We now briefly review the method that ref.~\cite{Manohar:2017eqh} used to compute the photon PDF, before applying the same procedure to the other PDFs. The computation of \fig{pdfop} gives (see sec.~6.1 of ref.~\cite{Manohar:2017eqh}, and dropping vacuum polarization corrections)
%%%
\begin{align}\label{eq:6.11}
 f_\gamma(x,\mu) 
&= \frac{8 \pi \alpha(\mu) \smu^{2\epsilon}}{x\, } \frac{1}{(4\pi)^{D/2}} \frac{1}{\Gamma(D/2-1)} 
 \int_x^1 \frac{\rd z}{z} \int_{\frac{m_p^2 x^2}{1-z}}^\infty \frac{\rd Q^2}{Q^2} P_\gamma 
 \nn
 & \quad \times  \big[Q^2(1-z)-x^2 m_p^2\big]^{D/2-2}
 \Big[-\frac{ z}{x (p^-)^2}\Big] \big[ (\bar n\cdot q)^2\, W^{(D) \lambda}_{\lambda}
+ q^2\, \bar n^\alpha \bar n^\beta  W^{(D)}_{\alpha \beta}\big]  \,,
\nn
f_{\Delta \gamma} (x,\mu) 
 &=  \frac{8 \pi \alpha(\mu) \smu^{2\epsilon} }{x\, } \frac{1}{(4\pi)^{D/2}} \frac{1}{\Gamma(D/2-1)}  \int_x^1 \frac{\rd z}{z}  \int_{\frac{m_p^2 x^2}{1-z}}^\infty  \frac{\rd Q^2}{Q^2} P_\gamma  \nn
 & \quad \times \big[Q^2(1-z)-x^2 m_p^2\big]^{D/2-2}  \Big[-\frac{ z}{\img x (p^-)^2}\Big] (\bar n \cdot q)\, \bar n^\alpha q^\beta \epsilon_{\alpha \mu \beta \nu} W^{(D)\mu \nu}\,.
\end{align}
%%%
Here
%%%
\bea
z \equiv \frac{x}{x_{\rm bj}}
\,,\eea 
%%%
$Q^2=-q^2$ is the momentum transfer, and $W_{\mu\nu}$ is evaluated at $(x_{\rm bj},Q^2)$. The label $D$ is a reminder that the hadronic tensor (and the couplings) are evaluated in $D=4-2\epsilon$ dimensions, and
%%%
\begin{align}  \label{eq:Sdef}
  {\cal S}^2=\frac{e^{\gamma_{\rm E}}}{4\pi}\,.
\end{align}
%%%
In \eq{6.11}, we included $P_\gamma=1$, as it will be replaced by other factors for the electroweak case, see \eq{2.22} through \eq{2.24} below. 
The $W_{\mu \nu}$ terms in \eq{6.11} can be written in terms of the structure functions using \eq{sf}, 
%%%
\begin{align} \label{eq:6.7}
 -\frac{ z}{x (p^-)^2} \big[ (\bar n\cdot q)^2\, W^{(D) \lambda}_{\lambda}
+ q^2\, \bar n^\alpha \bar n^\beta  W^{(D)}_{\alpha \beta}\big] &=  -z^2 F_{L, D}(x/z,Q^2) + \biggl(z p_{\gamma q}(z)+ \frac{2 m_p^2
      x^2}{Q^2} \biggr) \nn
      & \quad \times F_{2, D}(x/z,Q^2) 
      - 2 \epsilon\, z x\, F_{1, D}(x/z,Q^2) \,, \nn
  -\frac{ z}{i x (p^-)^2}  (\bar n \cdot q) \bar n^\alpha q^\beta \epsilon_{\alpha \mu \beta \nu} W^{(D)\mu \nu} &= -x \biggl(2-z -  \frac{2Q^2_{-2\epsilon}}{Q^2}\biggr){F}_{3,D}(x/z,Q^2)\,.
\end{align}
%%%
We retain the $F_3$ term, even though $F_3^{(\gamma)}=0$, since we will need it for the other PDFs. The splitting function in \eq{6.7} is
%%%
\begin{align}\label{eq:2.20a}
p_{\gamma q}(z) &= \frac{1+(1-z)^2}{z},
\end{align}
%%%
and
%%%
\begin{align}\label{eq:2.20}
Q^2_{-2\epsilon} &\equiv Q^2-Q^2_4
\end{align}
%%%
is the piece of $Q^2$ in fractional dimensions. Since we already averaged over the angular directions in obtaining \eq{6.11}, we can simply replace
%%%
\begin{align}\label{eq:2.21}
Q^2_{-2\epsilon} \to \frac{D-4}{D-2}\, Q_\perp^2  = \frac{D-4}{D-2} \left[ Q^2(1-z)- x^2 m_p^2\right].
\end{align}
%%%

We can now immediately get the other transverse PDFs. 
The only change is the replacement of the photon coupling and propagator by those for massive gauge bosons, and using the appropriate structure function. The $W^+$ PDF uses the $\bar \nu$ structure functions, and the replacement
%%%
\begin{align}\label{eq:2.22}
\alpha  &\to \frac{1}{8} \alpha_2,  & P_\gamma &\to P_W=\frac{Q^4}{(Q^2+M_W^2)^2}\,,
\end{align}
%%%
and the $W^-$ PDF uses \eq{2.22} with the $\nu$ structure functions. The $Z$ PDF has
%%%
\begin{align}\label{eq:2.23}
\alpha &\to \frac{1}{4} \alpha_Z, & P_\gamma &\to P_Z=\frac{Q^4}{(Q^2+M_Z^2)^2}\,.
\end{align}
%%%
The $\gamma Z$ and $Z \gamma$ PDFs use the $\gamma Z$ and $Z \gamma$ structure functions, with
%%%
\begin{align}\label{eq:2.24}
\alpha &\to \frac{1}{2} \sqrt{\alpha \alpha_Z} ,   & P_\gamma &\to P_{\gamma Z}=\frac{Q^2}{(Q^2+M_Z^2)}\,,
\end{align}
%%%
where $\alpha_2 = \alpha/\!\sin^2 \theta_W$ and $\alpha_Z= \alpha/(\sin^2 \theta_W \cos^2 \theta_W)$.

Proceeding as in ref.~\cite{Manohar:2017eqh}, the integral in \eq{6.11} over $Q^2$ is divided into an integral from $m_p^2 x^2/(1-z)$ to $\mu^2/(1-z)$, and from $\mu^2/(1-z)$ to $\infty$, where we assume $\mu \gg m_p$. Following the terminology of ref.~\cite{Manohar:2017eqh}, the two contributions are called the ``physical factorization'' term $f^{\text{PF}}$ and the \MSbar\ correction, $f^{\MSbar}$. The physical factorization integral is finite, so one can set $D=4$. The \MSbar\ integral is divergent, and needs to be evaluated in the \MSbar\ scheme (hence the name) to get the \MSbar\ PDF. As an example, we illustrate this for the $W^+$ PDF. Using eqs.~\eqref{eq:6.11}, \eqref{eq:6.7}, and \eqref{eq:2.22},
%%%
\begin{align}\label{eq:2.25}
x f^{\text{PF}}_{W^+_T}(x,\mu) 
&= \frac{\alpha_2(\mu) }{16 \pi  } 
 \int_x^1 \frac{\rd z}{z} \int_{\frac{m_p^2 x^2}{1-z}}^{\frac{\mu^2}{1-z}}  \frac{\rd Q^2}{Q^2} \frac{Q^4}{(Q^2+M_W^2)^2} \nn
 &\quad \times \bigg[ -z^2F^{(\bar \nu)}_{L}(x/z,Q^2) + \bigg(z p_{\gamma q}(z)+ \frac{2 m_p^2
      x^2}{Q^2} \bigg) F^{(\bar \nu)}_{2}(x/z,Q^2)  \bigg] 
\,, \end{align}
 %%%
and
%%%
\begin{align}\label{eq:2.26}
x f^{\MSbar}_{W^+_T}(x,\mu) 
&= \pi \alpha_2(\mu)\smu^{2\epsilon} \frac{1}{(4\pi)^{D/2}} \frac{1}{\Gamma(D/2-1)} 
 \int_x^1 \frac{\rd z}{z} \int_{\frac{\mu^2}{1-z}}^\infty  \frac{\rd Q^2}{Q^2} \frac{Q^4}{(Q^2+M_W^2)^2}\\
 &\quad \times[Q^2(1-z)]^{D/2-2} \left[ -z^2(1-\epsilon)  F^{(\bar \nu)}_{L, D}(x/z,Q^2) + \left(z p_{\gamma q}(z) - \epsilon z^2\right) F^{(\bar \nu)}_{2, D}(x/z,Q^2) \right].
\nonumber
 \end{align}
 %%%
Since the integral in \eq{2.26} is for $Q^2 \gg m_p^2$, we have dropped $m_p^2/Q^2$ terms. Changing variables to
%%%
\begin{align}\label{eq:6.17}
  s & = \frac{Q^2(1-z)}{\mu^2} \,,
\end{align}
%%%
gives
%%%
\begin{align}\label{eq:2.27}
x f^{\MSbar}_{W^+_T}(x,\mu) 
&= \frac{\alpha_2(\mu) }{16 \pi  } \frac{e^{\epsilon \gamma_E}}{\Gamma(1-\epsilon)} 
 \int_x^1 \frac{\rd z}{z} \int_1^\infty  \frac{\rd s}{s^{1+\epsilon}} \frac{\mu^4 s^2 }{\left[\mu^2 s+M_W^2(1-z)\right]^2}\\
 &\quad\times \left[ -z^2(1-\epsilon)  F^{(\bar \nu)}_{L, D}\bigl(x/z,\mu^2 s/(1-z)\bigr) + \left(z p_{\gamma q}(z) -  \epsilon z^2\right) F^{(\bar \nu)}_{2, D}\bigl(x/z,\mu^2 s/(1-z)\bigr)  \right] .
\nonumber \end{align}
%%%
Since $\mu$ is large, the dependence of $F_i(x/z,\mu^2 s/(1-z))$ on $\mu$ is perturbative. To lowest order in $\alpha_s$ and $\alpha_2$, we can therefore set the second argument of $F_i$ to $\mu^2$ without incurring large logarithms, and drop the $F_L$ term since it is order $\alpha_s$. This results in
%%%
\begin{align}\label{eq:2.28}
x f^{\MSbar}_{W^+_T}(x,\mu) 
&= \frac{\alpha_2(\mu) }{16 \pi  } \frac{e^{\epsilon \gamma_E}}{\Gamma(1-\epsilon)} 
 \int_x^1 \frac{\rd z}{z} \int_1^\infty  \frac{\rd s}{s^{1+\epsilon}} \frac{\mu^4 s^2 }{\left[\mu^2 s+M_W^2(1-z)\right]^2}
 \nn & \quad \times 
 \bigl[ \bigl(z p_{\gamma q}(z) -  \epsilon z^2\bigr) F^{(\bar \nu)}_{2, D}(x/z,\mu^2)  \bigr] \,.
 \end{align}
 %%%
 The $s$ integral yields
 %%%
\begin{align}\label{eq:24}
\frac{e^{\epsilon \gamma_E}}{\Gamma(1-\epsilon)} \int_{1}^\infty  \frac{\mathd s}{s^{1+\epsilon}} \frac{\mu^4 s^2}{[\mu^2 s+ M_W^2(1-z)]^2}
&=\frac{1}{\epsilon} + \ln \frac{\mu^2} {M_W^2(1-z)+\mu^2} -\frac{M_W^2(1-z)}{M_W^2(1-z)+\mu^2} + \mathcal{O}\left(\epsilon\right).
\end{align}
%%%
The $1/\epsilon$ term is cancelled by the UV counterterm, and the sum of \eqs{2.25}{2.28} gives
%%%
\begin{align} \label{eq:16.81}
x f_{W^+_T}(x,\mu) &=
  \frac{ \alpha_2(\mu) }{16\pi} \int_{x}^{1} 
  \frac{\df z}{z} \Biggl\{  \biggl[ \int^{\frac{\mu^2}{1-z}}_{\frac{m_p^2 x^2} {1-z}}
  \frac{\df Q^2}{Q^2} \frac{Q^4}{(Q^2+M_W^2)^2} \nn
& \quad \times \biggl(
    -z ^2 F^{(\bar \nu)}_L(x/z,Q^2)
  +\Bigl(
  z p_{\gamma q}(z)
    + \frac{2 x ^2 m_p^2}{Q^2}
   \Bigr) F^{(\bar \nu)}_2(x/z,Q^2) 
  \biggr)\biggr] \nn
& \quad  +
    z p_{\gamma q}(z)    \left( \ln \frac{\mu^2}{M_W^2(1-z)+\mu^2}  - \frac{M_W^2(1-z)}{M_W^2(1-z)+\mu^2}\right)F^{(\bar \nu)}_2(x/z,\mu^2)  \nn 
& \quad  -z^2  F^{(\bar \nu)}_2(x/z,\mu^2)  \biggr\}\ + \mathcal{O}( \alpha_2^2)\,.
\end{align}
%%%

The $1/\epsilon$ counterterm agrees with the anomalous dimensions for PDF evolution computed in refs.~\cite{Bauer:2017isx,Manohar:2018kfx}. Alternatively, one can also directly take the $\mu$ derivative of \eq{16.81}, for which the contribution  from the upper limit of the first integral cancels the contribution from rational function of $\mu^2$ and $M_W^2$, leaving the usual evolution
%%%
\begin{align}\label{eq:2.34}
\mu \frac{\rd}{\rd \mu} f_{W^+_T}(x,\mu) &= \frac{ \alpha_2(\mu) }{8\pi }  \int_{x}^{1} 
  \frac{\df z}{z}\, p_{\gamma q}(z)\, \frac{ F^{(\bar \nu)}_2(x/z,Q^2) }{x/z}
  + \mathcal{O}(\alpha_2\, \alpha_s) \,.
\end{align}
%%%
The largest effect not included is the QCD evolution of $ F^{(\bar \nu)}$.
Eq.~\eqref{eq:2.34} agrees with the anomalous dimension in refs.~\cite{Bauer:2017isx,Manohar:2018kfx},
%%%
\begin{align}\label{eq:2.35}
\mu \frac{\rd}{\rd \mu} f_{W^+_T}(x,\mu) 
&= \frac{ \alpha_2(\mu) }{2\pi} \int_{x}^{1} 
  \frac{\df z}{z}\,  p_{\gamma q}(z)  \bigl[f_{u_L}(x/z,Q^2) + f_{\bar d_R}(x/z,Q^2) \bigr] + \ldots
\end{align}
%%%
using \eq{2.17}.
In obtaining \eq{2.34} we can neglect $F_L$, the $\mu$-dependence of the structure functions and $\alpha_2(\mu)$, since these give terms that are higher order in $\alpha_2$ or $\alpha_s$. The diagonal $WW$ term in the PDF evolution, which contains Sudakov double logarithms, is also missing, since $f_{W_T}$ only starts at order $\alpha_2$.

Similarly, for $f_{\Delta W_T}$, using \eqs{6.11}{6.7},
%%%
\begin{align}
f^{\text{PF}}_{\Delta W_T^+} (x,\mu) 
 &=  -\frac{\alpha_2(\mu)}{ 16 \pi}  \int_x^1 \frac{\rd z}{z}  \int_{\frac{m_p^2 x^2}{1-z}}^{\frac{\mu^2}{1-z}}  \frac{\rd Q^2}{Q^2}  \frac{Q^4}{(Q^2+M_W^2)^2} \left(2-z \right){F}^{(\bar \nu)}_{3,D}(x/z,Q^2)
\end{align}
%%%
and
%%%
\begin{align}
f^{\MSbar}_{\Delta W_T^+} (x,\mu) 
 &=  -\frac{ \alpha_2(\mu)}{16 \pi} \frac{e^{\epsilon \gamma_E}}{\Gamma(1-\epsilon)}  \int_x^1 \frac{\rd z}{z} \int_1^\infty  \frac{\rd s}{s^{1+\epsilon}} \frac{\mu^4 s^2 }{\left[\mu^2 s+M_W^2(1-z)\right]^2}\nn
 & \quad \times  \Bigl(2-z + \frac{4\epsilon}{2-2\epsilon}(1-z) \Bigr){F}^{(\bar \nu)}_{3,D}\bigl(x/z,\mu^2 s/(1-z)\bigr) \,.
\end{align}
%%%
Replacing the second argument of $F_3$ by $\mu^2$, as before, and using \eq{24} gives
%%%
\begin{align} \label{eq:16.9}
f_{\Delta W_T^+}(x,\mu) &=
 - \frac{ \alpha_2(\mu) }{16\pi}   \int_{x}^{1} 
  \frac{\df z}{z}\, \biggl\{  \biggl[
  \int^{\frac{\mu^2}{1-z}}_{\frac{m_p^2 x^2} {1-z}}
  \frac{\df Q^2}{Q^2} \frac{Q^4}{(Q^2+M_W^2)^2}  \left(2-z\right) F^{(\bar \nu)}_3(x/z,Q^2) \biggr] \nn
& \quad   + \left(
    2-z \right)  \left( \ln \frac{\mu^2}{M_W^2(1-z)+\mu^2}  - \frac{M_W^2(1-z)}{M_W^2(1-z)+\mu^2}\right) F^{(\bar \nu)}_3(x/z,\mu^2) \nn
& \quad +  
  2(1-z)  F^{(\bar \nu)}_3(x/z,\mu^2) \biggr\} + \mathcal{O} ( \alpha_2^2)\,.
\end{align}
%%%
Differentiating \eq{16.9} w.r.t.\ $\mu$ gives the evolution equation
%%%
\begin{align}\label{eq:2.30}
\mu \frac{\rd}{\rd \mu} f_{\Delta W^+_T}(x,\mu) &= -\frac{ \alpha_2(\mu) }{8\pi }  \int_{x}^{1} 
  \frac{\df z}{z}\, (2-z)  F^{(\bar \nu)}_3(x/z,Q^2) 
  + \mathcal{O}(\alpha_2\, \alpha_s) \,,
\end{align}
which agrees with ref.~\cite{Manohar:2018kfx}. 

The $W^-_T$ PDF is given by \eq{16.81}, \eq{16.9} with  $\bar \nu$ structure functions replaced by $\nu$ structure functions. The $Z_T$ PDF is given by \eq{16.81}, \eq{16.9} with  $\bar \nu$ structure functions replaced by $Z$ structure functions, $M_W \to M_Z$, and $\alpha_2 \to 2 \alpha_Z$. The $\gamma Z$ PDFs require the $s$ integral
%%%
\begin{align}\label{eq:24Z}
\frac{e^{\epsilon \gamma_E}}{\Gamma(1-\epsilon)} \int_{1}^\infty  \frac{\mathd s}{s^{1+\epsilon}} \frac{\mu^2 s}{\mu^2 s+ M_Z^2(1-z)}
&=\frac{1}{\epsilon} + \ln \frac{\mu^2} {\mu^2+M_Z^2(1-z)} + \mathcal{O}\left(\epsilon\right),
\end{align}
%%%
since there is only one massive propagator. This gives
%%%
\begin{align} \label{eq:16.8}
x f_{\gamma Z_T}(x,\mu) &=
  \frac{ \sqrt{\alpha(\mu) \alpha_Z(\mu)} }{4\pi}   \int_{x}^{1} 
  \frac{\df z}{z} \biggl\{ \biggl[
  \int^{\frac{\mu^2}{1-z}}_{\frac{m_p^2 x^2} {1-z}}
  \frac{\df Q^2}{Q^2} \frac{Q^2}{Q^2+M_Z^2} \\
&\quad\times \biggl(
    -z ^2F^{(\gamma Z )}_L(x/z,Q^2)
  +\Bigl(
  z p_{\gamma q}(z)    + \frac{2 x ^2 m_p^2}{Q^2}
   \Bigr) F^{(\gamma Z )}_2(x/z,Q^2) 
  \biggr) \biggr] \nn
& \quad  +
  z p_{\gamma q}(z) \left( \ln \frac{\mu^2}{M_Z^2(1-z)+\mu^2} \right)  F^{(\gamma Z)}_2(x/z,\mu^2)    -z^2  F^{(\gamma Z)}_2(x/z,\mu^2)   \biggr\}
  + \mathcal{O}(\alpha_2^2)\,,
\nonumber \end{align}
%%%
where we consider both $\alpha$ and $\alpha_Z$ of order $\alpha_2$  in writing $\mathcal{O}(\alpha_2^2)$.
Similarly,
%%%
\begin{align} \label{eq:16.10}
f_{\Delta \gamma Z_T}(x,\mu) &=
 -  \frac{ \sqrt{\alpha(\mu) \alpha_Z(\mu)} }{4\pi}   \int_{x}^{1} 
  \frac{\df z}{z} \Biggl\{ \biggl[
  \int^{\frac{\mu^2}{1-z}}_{\frac{m_p^2 x^2} {1-z}}
  \frac{\df Q^2}{Q^2} \frac{Q^2}{Q^2+M_Z^2}  \left(2-z\right) F^{(\gamma Z)}_3(x/z,Q^2) \biggr]  \nn
& \quad +  \left(
    2-z \right)   \left( \ln \frac{\mu^2}{M_Z^2(1-z)+\mu^2} \right) F^{(\gamma Z )}_3(x/z,\mu^2) + 
  2(1-z)  F^{(\gamma Z)}_3(x/z,\mu^2)  \biggr\} 
  \nn & \quad
  + \mathcal{O} ( \alpha_2^2)
\,,\end{align}
%%%
and the $Z\gamma$ PDF is obtained by $\gamma Z \to Z \gamma$.

%%%%%%%%%%%%%%%%%%%%%%%%%%%%%%%%%%%%%%%%%%%%%%%%%%%%%%%%%%%%%%%%%%%%%%%%%%%%%%%%
\section{Longitudinal gauge boson PDFs}
\label{sec:longitudinal}
%%%%%%%%%%%%%%%%%%%%%%%%%%%%%%%%%%%%%%%%%%%%%%%%%%%%%%%%%%%%%%%%%%%%%%%%%%%%%%%%

In this section we repeat the analysis of \sec{transverse} for the PDFs of longitudinal gauge bosons. We start again with defining them, using the equivalence theorem to express them in terms of scalar PDFs, and then calculate them in terms of structure functions.

%===============================================================================
\subsection{Definition}
%===============================================================================

The operators in \eq{W} give the PDFs for transversely polarized gauge bosons. Longitudinally polarized gauge bosons are not produced at leading power in $M/Q$ by the gauge field-strength tensor. Instead, they have to be computed in terms of Goldstone bosons using the Goldstone-boson equivalence theorem~\cite{chanowitz,bohmbook}, as was done for electroweak corrections to scattering amplitudes in refs.~\cite{Chiu:2009ft,Chiu:2009mg}. The scalar (Higgs) PDFs we need are
%%%
\begin{align}\label{eq:higgs}
O_{H^+}(r^-) &= \frac{r^- }{ 2 \pi} \int_{-\infty}^\infty \! \df \xi\, e^{-\img \xi r^-}
[ H^\dagger (\bar n \xi )\, \W (\bar n \xi )]_1\ [\W^\dagger (0) \,  H(0 ) ]^1 \,, \nn 
O_{ \bar H^-}(r^-) &= \frac{r^- }{ 2 \pi} \int_{-\infty}^\infty\! \df \xi\, e^{-\img \xi r^-}
[\W^\dagger (\bar n \xi) \,  H(\bar n \xi ) ]^1\ [ H^\dagger (0 )\, \W (0)]_1\,, \nn 
O_{H^0}(r^-) &= \frac{r^- }{ 2 \pi} \int_{-\infty}^\infty\! \df \xi\, e^{-\img \xi r^-}
[ H^\dagger (\bar n \xi )\, \W (\bar n \xi )]_2\ [\W^\dagger (0) \,  H(0 ) ]^2 \,, \nn 
O_{\bar H^0}(r^-) &= \frac{r^- }{ 2 \pi} \int_{-\infty}^\infty\! \df \xi\, e^{-\img \xi r^-}
[\W^\dagger (\bar n \xi) \,  H(\bar n \xi ) ]^2\ [ H^\dagger (0 )\, \W (0)]_2\,, \nn 
O_{\bar H^0 H^0}(r^-) &= \frac{r^- }{ 2 \pi} \int_{-\infty}^\infty\! \df \xi\, e^{-\img \xi r^-}
[\W^\dagger (\bar n \xi) \,  H(\bar n \xi ) ]^2 \ [\W^\dagger (0) \,  H(0 ) ]^2 \,, \nn 
O_{H^0 \bar H^0}(r^-) &= \frac{r^- }{ 2 \pi} \int_{-\infty}^\infty\! \df \xi\, e^{-\img \xi r^-}
 [ H^\dagger (\bar n \xi )\, \W (\bar n \xi )]_2 \  [ H^\dagger (0 )\, \W (0 )]_2 \,,
\end{align}
%%%
with $SU(2) \times U(1)$ Wilson lines $\W$. The indices $1,2$ in \eq{higgs} pick out the charged and neutral components of the Higgs multiplet, 
%%%
\begin{align}\label{eq:3.2}
H &= 
\begin{pmatrix}
H^+  \\
H^0 \end{pmatrix} 
=
\frac{1}{\sqrt 2}\! \begin{pmatrix}
 \img \sqrt2\, \varphi^+ \\
v + h - \img \varphi^3 \end{pmatrix}\,,
\end{align}
%%%
in the unbroken and broken phase, respectively. Here $h$ is the physical Higgs particle, and the unphysical scalars $\varphi^3,\varphi^\pm=(\varphi^1 \mp \img \varphi^2)/\sqrt{2}$ are related to the longitudinal gauge bosons $Z_L, W^\pm_L$ through the Goldstone-boson equivalence theorem. For the incoming gauge bosons this is given by $\img \varphi^+ \to W^+_L$, $\img \varphi^3 \to Z_L$, and for gauge bosons on the other side of the cut in \fig{pdfop}, $-\img \varphi^+ \to W^+_L$, $-\img \varphi^3 \to Z_L$. This leads to
%%%
\begin{align}\label{eq:L}
f_{W^+_L}(x,\mu) &= \braket{p | O_{H^+}(x p^-) | p},  \nn 
f_{W^-_L}(x,\mu) &= \braket{p | O_{H^-}(x p^-) | p},  \nn 
f_{Z_L}(x,\mu) &= \tfrac12  \braket{p | \left[ O_{H^0}(x p^-) + O_{\bar H^0}(x p^-) - O_{\bar H^0 H^0}(x p^-) - O_{H^0 \bar H^0}(x p^-) \right] | p},  \nn 
f_{h}(x,\mu) &= \tfrac12  \braket{p | \left[ O_{H^0}(x p^-) + O_{\bar H^0}(x p^-) + O_{\bar H^0 H^0}(x p^-)  + O_{H^0 \bar H^0}(x p^-) \right] | p} , \nn 
f_{hZ_L}(x,\mu) &= \tfrac12  \braket{p | \left[ O_{H^0}(x p^-) - O_{\bar H^0}(x p^-) + O_{\bar H^0 H^0}(x p^-) - O_{H^0 \bar H^0}(x p^-) \right] | p},  \nn 
f_{Z_L h}(x,\mu) &= \tfrac12  \braket{p | \left[  O_{H^0}(x p^-) - O_{\bar H^0}(x p^-) - O_{\bar H^0 H^0}(x p^-) + O_{H^0 \bar H^0}(x p^-) \right] | p}, 
\end{align}
%%%
which are equivalent to eqs.~(5.4) and (5.5) of ref.~\cite{Manohar:2018kfx}. Note that $f_{h Z_L}$ and $f_{Z_L h}$ are complex conjugates of each other.

It is instructive to rederive \eq{L}, by expanding the Wilson lines in eq.~\eqref{eq:higgs}  to first order
%%%
\begin{align}\label{eq:3.3}
H^\dagger (x)\W(x)  &= H^\dagger(x)\, P \exp \left\{- \img \int^0_{\infty} \rd s\ \bar n \cdot \left[g_2 W(x + \bar n s) + g_1 B(x+\bar n s)\right]\right\}  \nn
&=\begin{pmatrix}
- \img \, \varphi^{-}(x) &
\quad \frac{1}{\sqrt 2}\left[ {v + h(x) + \img \varphi^3(x)} \right]\end{pmatrix}  \nn
&\quad - \img  \frac{v}{\sqrt 2} \int^0_{\infty} \rd s \begin{pmatrix}
\frac{g_2}{\sqrt 2}\, \bar n \cdot  W^-(x+\bar n s)  &\quad  -\frac12 g_Z\, \bar n \cdot  Z(x+\bar n s)
 \end{pmatrix} + \ldots
,\end{align} 
%%%
where $g_2 = e/\!\sin \theta_W$ and $g_Z = e/(\sin \theta_W \cos \theta_W)$.
Using integration by parts, we have the identity 
%%%
\begin{align}\label{eq:3.5}
& -\img \,r^- \int_{-\infty}^\infty \rd \xi\ e^{- \img \xi r^-}\int^0_{\infty} \rd s\ \bar n \cdot Z(\bar n \xi + \bar n s)
=  \int_{-\infty}^\infty \rd \xi  \left[ \frac{\rd}{\rd \xi} e^{- \img \xi r^-}\right] \int^0_{\infty} \rd s\ \bar n \cdot Z(\bar n \xi + \bar n s) \nn
&= -\int_{-\infty}^\infty \rd \xi\  e^{- \img \xi r^-} \frac{\rd}{\rd \xi}\int^0_{\infty} \rd s\ \bar n \cdot Z(\bar n \xi + \bar n s) 
= -\int_{-\infty}^\infty \rd \xi\  e^{- \img \xi r^-} \bar n \cdot Z(\bar n \xi),
\end{align}
%%%
setting the gauge field at infinity to zero, and similarly for the $W^-$ term. 
As a result, 
%%%
\begin{align}\label{eq:3.7}
 H^\dagger(\bar n \xi) \W(\bar n \xi) &= \begin{pmatrix}
  - \img \varphi^-(\bar n \xi) - \frac{M_W}{r^-}\bar n \cdot W^- (\bar n \xi)    &\quad \frac{1}{\sqrt 2}\left[
v + h(\bar n \xi) +   \img \varphi^3(\bar n \xi) + \frac{M_Z}{r^-} \bar n \cdot Z(\bar n \xi) \right] \end{pmatrix}
\nn & \quad  + \ldots
\end{align} 
%%%
inside an integral of the form as in \eq{higgs},
where we used $M_W=g_2 v/2$ and $M_Z =g_Z v/2$.  One can make a similar substitution for the $\W^\dagger(0) H(0)$ term. The argument $0$ does not depend on the integration variable $\xi$. However, we can use translation invariance of \eq{higgs} to switch the field arguments in \eq{higgs} from $\bar n \xi$ and $0$ to $0$ and $-\bar n \xi$. Eq.~(\ref{eq:3.5}) can be applied again, with an additional minus sign because of the switch $\bar n \xi \to - \bar n \xi$. Then $\W^\dagger(0) H(0)$ is given by the conjugate of \eq{3.7} with $r^- \to - r^-$ and $\xi \to 0$. 

The linear combinations in \eq{3.7} are those required by the equivalence theorem. Exploiting gauge invariance, we can evalute the PDFs in the broken phase  in unitary gauge using \eq{3.7} with $\varphi_i \to 0$. This does not affect renormalizability, since the PDFs in the broken phase only have radiative corrections due to dynamical gluons and photons, i.e.~the $W$ and $Z$ are treated as static fields in the same way as heavy quark fields in heavy quark effective theory.
Thus we reproduce \eq{L}, identifying the longitudinal PDFs as
%%%
\begin{align}\label{eq:L2}
f_{W^+_L}(x,\mu) &= \frac{M_W^2}{ 2 \pi r^-} \int_{-\infty}^\infty\! \df \xi\, e^{-\img  \xi r^-} \braket{p |
[ \bar n \cdot W^-(\bar n \xi )\, \W (\bar n \xi )]\ [\W^\dagger (0) \, \bar n \cdot W^+(0) ] | p},  \nn 
f_{W^-_L}(x,\mu) &= \frac{M_W^2}{ 2 \pi r^-} \int_{-\infty}^\infty\! \df \xi\, e^{-\img  \xi r^-} \braket{p |
[ \bar n \cdot W^+(\bar n \xi )\, \W (\bar n \xi )]\ [\W^\dagger (0) \, \bar n \cdot W^-(0) ] | p},  \nn 
f_{Z_L}(x,\mu) &= \frac{M_Z^2}{ 2 \pi r^-} \int_{-\infty}^\infty\! \df \xi\, e^{-\img  \xi r^-} \braket{p |
 \bar n \cdot Z(\bar n \xi)\ \bar n \cdot Z(0)  | p},  \nn 
f_{h}(x,\mu) &=  \frac{r^-}{ 2 \pi } \int_{-\infty}^\infty\! \df \xi\, e^{-\img  \xi r^-} \braket{p |
h(\bar n \xi)\ h(0)  | p}, \nn
f_{hZ_L}(x,\mu) &= \frac{M_Z}{ 2 \pi } \int_{-\infty}^\infty\! \df \xi\, e^{-\img  \xi r^-} \braket{p |
h (\bar n \xi) \ \bar n \cdot Z(0) | p} ,  \nn
f_{Z_L h}(x,\mu) &= \frac{M_Z}{ 2 \pi } \int_{-\infty}^\infty\! \df \xi\, e^{-\img  \xi r^-} \braket{p |
 \bar n \cdot Z(\bar n \xi)\ h(0) | p} \,,
\end{align}
%%%
where the Wilson lines $\W$ in the $W$ PDFs only contain photon fields.

%===============================================================================
\subsection{Evaluation}
%===============================================================================

Before evaluating the longitudinal gauge boson PDFs, we note that the Higgs PDFs in \eq{L} are suppressed,
%%%
\begin{align}\label{eq:L3}
f_{h}(x,\mu) &= \mathcal{O} \bigg( \frac{m_p^2}{M_Z^2} \bigg), &
f_{hZ_L}(x,\mu) &= \mathcal{O} \bigg( \frac{m_p}{M_Z} \bigg),  &
f_{Z_L h}(x,\mu) & = \mathcal{O} \bigg( \frac{m_p}{M_Z} \bigg)
\,.\end{align}
%%%
This happens because the gauge field couplings to the proton are of order $g_2,g_Z$, whereas the dominant coupling of the Higgs field to the proton is given by the scale anomaly~\cite{Voloshin:1980zf}, and is order $m_p/v$ (for a pedagogical discussion see ref.~\cite{Chivukula:1989ds}). (There are of course also contributions of the order of light fermion Yukawa couplings $m_{u,d}/v$.) We therefore neglect the Higgs PDFs in \eq{L3} in this paper, but they can be computed using the same method as the gauge boson PDFs.

We now repeat the steps in \sec{eva_T} for longitudinal gauge bosons, starting with $f_{W^+_L}$.
The matrix element of the PDF operator gives 
%%%
\begin{align}\label{eq:3.10}
 f_{W^+_L} (x,\mu) 
&=   \frac{\pi  \alpha_2(\mu) \smu^{2\epsilon} }{x} \frac{1}{(4\pi)^{D/2}} \frac{1}{\Gamma(D/2-1)} 
 \int_x^1 \frac{\rd z}{z} \int_{\frac{m_p^2 x^2}{1-z}}^\infty  \frac{\rd Q^2}{Q^2} \frac{Q^4}{(Q^2+M_W^2)^2}\nn
 &\quad\times \big[Q^2(1-z)-x^2 m_p^2\big]^{D/2-2}
 \left[\frac{ z M_W^2}{x (p^-)^2}\right]  \bar n^\mu \bar n^\nu W^{(\bar \nu)}_{\mu \nu }   \,,
\end{align}
%%%
where 
%%%
\begin{align}
 \left[\frac{ z Q^2}{2x (p^-)^2}\right]  \bar n^\mu \bar n^\nu W^{(\bar \nu)}_{\mu \nu } &= \Big(  1-z - \frac{m_p^2 x^2}{Q^2} \Big)F^{(\bar \nu)}_2 + \frac14  z^2 F^{(\bar \nu)}_L \,,
\label{5}
\end{align}
%%%
which combine to yield 
%%%
\begin{align}\label{eq:3.11}
 f_{W^+_L} (x,\mu) 
&=   \frac{\pi  \alpha_2(\mu) \smu^{2\epsilon} }{x} \frac{1}{(4\pi)^{D/2}} \frac{1}{\Gamma(D/2-1)} 
 \int_x^1 \frac{\rd z}{z} \int_{\frac{m_p^2 x^2}{1-z}}^\infty  \frac{\rd Q^2}{Q^2}  \frac{Q^4}{(Q^2+M_W^2)^2}\\
 &\quad\times \big[Q^2(1-z)-x^2 m_p^2\big]^{D/2-2}
 \left[\frac{2 M_W^2}{Q^2 }\right]   \bigg[ \Big(  1-z - \frac{m_p^2 x^2}{Q^2} \Big)F^{(\bar \nu)}_2 + \frac14  z^2 F^{(\bar \nu)}_L  \bigg]\,. \nonumber
\end{align}
%%%
Using the same procedure of splitting the integral as before gives 
%%%
\begin{align}
  \label{eq:47}
x f_{W^+_L}(x,\mu) &= \frac{\alpha_2(\mu)}{8\pi }
\biggl\{
  \int_x^1 
  \frac{\rd z}{z} \biggl[
  \int^{\frac{\mu^2}{1-z}}_{\frac{m_p^2 x^2} {1-z}}
  \frac{\rd Q^2}{Q^2} \frac{M_W^2 Q^2}{(Q^2+M_W^2)^2}  \nn 
  & \quad \times
   \biggl(
  \Bigl(
1-z - \frac{x ^2 m_p^2}{Q^2} \Bigr) F^{(\bar \nu)}_2 (x/z,Q^2) 
  +
    \frac14 z ^2   F^{(\bar \nu)}_L (x/z,Q^2) 
  \biggr) \nn
& +\frac{M_W^2 (1-z)^2}{\mu^2 + M_W^2 (1-z)} F^{(\bar \nu)}_2 (x/z,\mu^2) 
 \biggr] \biggr\} + \mathcal{O} (\alpha_2^2)\,,
\end{align} 
%%%
and similarly for $f_{W^-_L}$, with the replacement $F^{(\bar \nu)}_i \to F^{( \nu)}_i$. For $f_{Z_L}$ this requires replacing $F^{(\bar \nu)}_i \to F^{(Z)}_i$, $M_W \to M_Z$ and $\alpha_2 \to 2 \alpha_Z$. 

Comparing \eq{47} with \eq{16.81} for the transverse $W$ PDF, we see that $W_L$ has an extra $M_W^2/Q^2$ factor in the $Q^2$ integral. The $W_T$ integral grows as $\ln \mu^2$ for large values of $\mu$, whereas the $W_L$ integral is finite, and dominated by $Q^2 \sim M_W^2$. The $W_L$ PDF is thus smaller than the $W_T$ PDF by $\ln \mu^2/M_W^2$. The split of the longitudinal PDFs into two pieces in \eq{47} is not necessary, and one can instead use \eq{47} with $\mu \to \infty$, but it is useful when comparing with the other PDFs. Differentiating \eq{47} with respect to $\mu$ gives
%%%
\begin{align}
  \label{eq:48}
\mu \frac{\rd}{\rd \mu}x f_{W^+_L}(x,\mu) &= 0 + \mathcal{O} ( \alpha_2  \alpha_s )\,,
\end{align} 
%%%
as expected.
Eq.~(\ref{eq:47}) was obtained starting from \eq{L2} in the broken phase. For $\mu$ much larger than $M_W$, we need the PDFs in the unbroken theory, 
which are related to Higgs PDFs by \eq{L}. Since there is no quark contribution to the Higgs PDF evolution when fermion Yukawa couplings are neglected, \eq{48} is expected.

%%%%%%%%%%%%%%%%%%%%%%%%%%%%%%%%%%%%%%%%%%%%%%%%%%%%%%%%%%%%%%%%%%%%%%%%%%%%%%%%
\section{Comparison with previous results}\label{sec:compare}
%%%%%%%%%%%%%%%%%%%%%%%%%%%%%%%%%%%%%%%%%%%%%%%%%%%%%%%%%%%%%%%%%%%%%%%%%%%%%%%%

$W$ and $Z$ PDFs have been computed previously using the effective $W,Z$ approximation~\cite{Chanowitz:1984ne,Dawson:1984gx,Kane:1984bb}, i.e.\ the  Fermi-Weizs\"acker-Williams~\cite{Fermi:1924tc,vonWeizsacker:1934nji,Williams:1934ad} approximation for electroweak gauge bosons. We will compare these with our results.

Considering for concreteness the transverse $W$ PDF, the leading-logarithmic contribution from \eq{16.81} is given by
%%%
\begin{align} \label{eq:3.15}
x f_{W^+_T}(x,\mu) &\approx
  \frac{ \alpha_2(\mu) }{16\pi} \  \int_{x}^{1} 
  \frac{\df z}{z} \int^{\mu^2}
  \frac{\df Q^2}{Q^2} \frac{Q^4}{(Q^2+M_W^2)^2} z p_{\gamma q}(z) F^{(\bar \nu)}_2(x/z,Q^2) \nn
   &\approx
  \frac{ \alpha_2(\mu) }{16\pi} \ln \frac{\mu^2}{M_W^2}  \int_{x}^{1} 
  \frac{\df z}{z} z p_{\gamma q}(z) F^{(\bar \nu)}_2(x/z,\mu^2) \,,
\end{align}
%%%
and agrees with the effective $W$ approximation result in refs.~\cite{Chanowitz:1984ne,Dawson:1984gx,Kane:1984bb}. The subleading terms (the last two lines in \eq{16.81}) are smaller by a factor of $\ln \mu^2/M_W^2$. These differ from the corresponding terms in previous results.

The longitudinal $W$ PDF is smaller by a factor of $\ln \mu^2/M_W^2$, and is obtained by integrating
%%%
\begin{align}
  \label{eq:49}
x f_{W^+_L}(x,\mu) &\approx \frac{\alpha_2(\mu)}{8\pi }
  \int_x^1 
  \frac{\rd z}{z} 
  \int_0^\infty
  \frac{\rd Q^2}{Q^2}  \frac{M_W^2 Q^2}{(Q^2+M_W^2)^2}
  \left(1-z \right) F^{(\bar \nu)}_2 (x/z,Q^2) \nn
  &\approx \frac{\alpha_2(\mu^2)}{8\pi }
  \int_x^1 
  \frac{\rd z}{z} 
  \left(1-z \right) F^{(\bar \nu)}_2 (x/z,\mu^2) \,,
\end{align} 
%%%
which agrees with refs.~\cite{Chanowitz:1984ne,Dawson:1984gx,Kane:1984bb}. Again, the subleading terms given by the last line in \eq{47} differ from previous results.

%%%%%%%%%%%%%%%%%%%%%%%%%%%%%%%%%%%%%%%%%%%%%%%%%%%%%%%%%%%%%%%%%%%%%%%%%%%%%%%%
\section{PDF computation using factorization methods}
\label{sec:factorization}
%%%%%%%%%%%%%%%%%%%%%%%%%%%%%%%%%%%%%%%%%%%%%%%%%%%%%%%%%%%%%%%%%%%%%%%%%%%%%%%%

In this section we present an alternative derivation for the massive gauge boson PDF using standard factorization methods. We will consider both transverse and longitudinal polarizations, and consider a massive gauge boson in a broken $U(1)$ theory to keep the presentation simple. Our calculation exploits the fact that the cross section for the hypothetical process of electron-proton scattering producing a new heavy lepton or scalar in the final state can be written in two ways: in terms of proton structure functions or  using proton PDFs. This approach was used in ref.~\cite{Manohar:2016nzj} for the photon case. The new feature in the calculation is broken gauge symmetry, which results in massive gauge bosons that can have a longitudinal polarization.

%===============================================================================
\subsection{Transverse polarization}
%===============================================================================

\begin{figure}
\centering
\includegraphics[width=6cm]{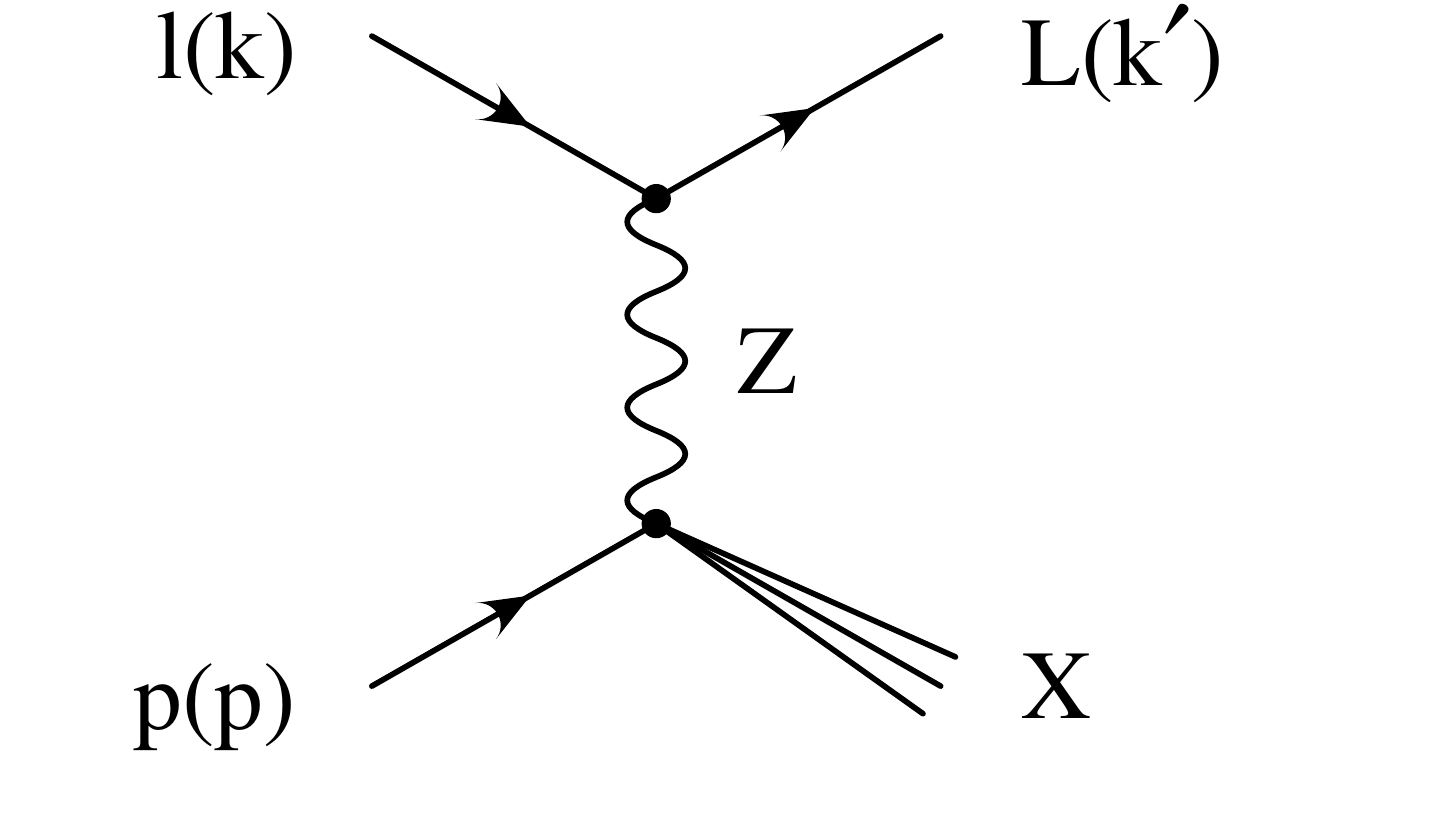}
\caption{Feynman diagram for heavy lepton production\ $l+p\rightarrow L+X$.}
\label{fig:5}
\end{figure}

 Following refs.~\cite{Manohar:2016nzj,Manohar:2017eqh},  consider the hypothetical inclusive scattering process
%%%
\bea\label{pr1}
l(k) + p(p) \rightarrow L(k') + X ,
\eea
%%%
shown in Fig.~\ref{fig:5}, where $l$ is a massless fermion, and $L$ is a fermion with mass $M_L$. We will assume that they interact with the massive $U(1)$ gauge boson (called $Z$) via a magnetic momentum coupling,
%%%
\begin{align}
\mathcal{L}_{\rm int} &= \frac{g}{\Lambda} \bar{L} \,\sigma^{\mu\nu} Z_{\mu\nu} \,l  + \text{h.c.} \,.
\end{align}
%%%
Here $g$ is the gauge coupling, and we work to leading order in the scale of the new interaction $\Lambda \gg v$. The interaction between $Z$ and the protons is governed by  $\cL_{Z p} = - g Z^\mu j_\mu$. We will now calculate the $Z$ PDF by first factorizing the cross section into the hadronic and leptonic tensor, and then factorizing it in terms of PDFs. In doing so, we assume $M_L \to \infty$.

The cross section averaged over initial spins and summed over final states is given by
%%%
\begin{align} \label{eq:4.4}
 \sigmaprobe{p} &= \frac{1}{4 p\cdot k} \int \frac{\mathd^4
    q}{(2\pi)^4} \frac{g^2}{(q^2-M_Z^2)^2}\ 4\pi W^{\mu\nu}(p,q) \;
  L_{\mu\nu}(k,q) \ 2\pi \delta[(k-q)^2-M_L^2]\, \theta(k^0-q^0)\nn
  &\quad \times \theta[(p+q)^2-m_p^2]\ \theta(p^0+q^0)\,,
\end{align}
%%%
in terms of the hadronic tensor $W_{\mu \nu}(p,q)$ and the leptonic tensor $L_{\mu \nu}(k,q)$, with $q=k-k'$. The lepton tensor, averaging over initial spins  and summing over final ones is
%%%
\begin{align}\label{eq:5.4}
L^{\mu\nu} &= \frac{1}{2} \frac{g^2}{\Lambda^2}\, {\rm tr}\left(\slashed{k}\, [\slashed{q},\gamma^\mu] (\slashed{k}'+M_L)[\gamma^\nu,\slashed{q}]\right) \nn
&=\frac{8g^2}{\Lambda^2}\Big[ (k\cdot q)\,(q^2 - 2k\cdot q) g^{\mu\nu}-2q^2k^\mu k^\nu+  (k\cdot q) \,(2q^\mu k^\nu + 2 q^\nu k^\mu - q^\mu q^\nu) \Big]\nn
&= \frac{4g^2}{\Lambda^2} \Bigl[ \left(M_L^2+Q^2\right) \left( q^\mu q^\nu  - M_L^2 g^{\mu \nu} \right) +4 Q^2 k^\mu k^\nu -2(M_L^2+Q^2) \left(k^\mu q^\nu+k^\nu q^\mu\right) \Bigr]\,.
\end{align}
%%%
The decomposition of the hadronic tensor in terms of structure functions was given by \eq{sf}. $F_3$ does not contribute to the spin-averaged cross section, since \eq{5.4} is symmetric in $\mu \leftrightarrow \nu$. The rest of the derivation of $\sigma_{lp}$ is then identical to sec. 3 of ref.~\cite{Manohar:2017eqh}, after accounting for the gauge boson mass in the propagator $1/q^4 \to 1/(q^2-M_Z^2)^2$. The total cross section in the $M_L \to \infty$ limit is thus given by
%%%
\begin{align}
\sigma_{lp} &=  \frac{g^4}{2 \pi \Lambda^2} \int_\xph^1 \frac{\rd z}{z}
  \int^{\frac{\mu^2}{1-z}}_{\frac{m_p^2 x^2}{1-z}} \frac{\df Q^2}{Q^2} \frac{Q^4}{(Q^2+M_Z^2)^2} \bigg[ -z ^2\, F_L(x/z,Q^2) +\biggl(
    z p_{\gamma q}(z) + \frac{2 x ^2 m_p^2}{Q^2} \biggr)
    F_2(x/z,Q^2) \bigg]\, \nn
    & \quad +  \frac{g^4}{2 \pi \Lambda^2} \int_\xph^1 \frac{\rd z}{z}
  \int_{\frac{\mu^2}{1-z}}^{\frac{M_L^2 (1-z)}{z}} \frac{\df Q^2}{Q^2} \frac{Q^4}{(Q^2+M_Z^2)^2} \bigg[ \biggl(
  -\frac{2 zQ^2}{M_L^2}+\frac{z^2 Q^2}{M_L^2} + z p_{\gamma q}(z) \biggr)
    F_2(x/z,Q^2) \bigg],
\end{align}
%%%
where the kinematic variables are 
%%%
\begin{align} \label{eq:5.6}
x_{\rm bj} = \frac{Q^2}{2 p \cdot q} &= \frac{x}{z}, & p\cdot k &= \frac{M^2}{2x}, &  s &=(p+k)^2 ,
\end{align}
%%%
and we have split the $Q^2$ integral into two parts. If $\mu$ is large compared to $\Lambda_{\rm QCD}$, we can replace $F_2(x/z,Q^2)$ in the second integral by $F_2(x/z,\mu^2)$ since the $\mu$ evolution is perturbative, and evaluate the integral to obtain
%%%
\begin{align} \label{eq:5.7}
\sigma_{lp} &=  \frac{g^4}{2 \pi \Lambda^2} \int_\xph^1 \frac{\rd z}{z}
  \int^{\frac{\mu^2}{1-z}}_{\frac{m_p^2 x^2}{1-z}} \frac{\df Q^2}{Q^2} \frac{Q^4}{(Q^2+M_Z^2)^2} \Bigg[ -z ^2\, F_L(x/z,Q^2) +\biggl(
    z p_{\gamma q}(z) + \frac{2 x ^2 m_p^2}{Q^2} \biggr)
    F_2(x/z,Q^2) \Bigg] \nn 
    & \quad +  \frac{g^4}{2 \pi \Lambda^2} \int_\xph^1 \frac{\rd z}{z} z p_{\gamma q}(z) \left[ 
    \ln \frac{ M_L^2(1-z)^2}{z[M_Z^2(1-z)+\mu^2]} - \frac{M_Z^2(1-z)}{M_Z^2(1-z)+\mu^2}\right] F_2(x/z,\mu^2) \nn
    & \quad +  \frac{g^4}{2 \pi \Lambda^2} \int_\xph^1 \frac{\rd z}{z} (-z^2+3z-2)
    F_2(x/z,\mu^2) \,.
\end{align}
%%%

We now factorize the cross section into a convolution of hard-scattering cross sections and parton distributions, 
%%%
\begin{align} \label{eq:5.8}
\sigma_{lp} (xs) =\sum_{a=Z,q,\ldots} \int_x^1 \frac{\df z}{z}\, \hat{\sigma}_{la}(z s,\mu)\, \frac{x}{z} f_{a/p} \left(\frac{x}{z}, \mu\right).
\end{align}
%%%
These hard-scattering cross sections are the same as in ref.~\cite{Manohar:2017eqh},
%%%
\begin{align}\label{eq:5.9}
\sigma_0 &= \frac{4 \pi g^2}{\Lambda^2},\nn
\hat \sigma_{lZ} &= \sigma_0 \, \delta(1 - z), \nn
\hat \sigma_{lq} &= \sigma_0 \frac{g^2}{8\pi^2} \bigg[-2 + 3z + z p_{\gamma q}(z) \ln \frac{M_L^2(1-z)^2}{z \mu^2}\bigg],
\end{align}
%%%
with $z=M_L^2/\hat s$. It should not come as a surprise that this only describes transverse polarizations, since it is the same as for photons. Indeed, the contribution to the cross section of this process is power suppressed by $M_Z^2/M_L^2$ for longitudinally polarized gauge bosons. Thus the factorization formula in \eq{5.8} gives the $Z$ PDF summed over the two transverse polarizations only.

We can then extract $f_{Z_T}$ by combining \eqs{5.7}{5.8},
%%%
\begin{align}\label{eq:5.10}
xf_{Z_T}(x) &=  \frac{g^2}{8 \pi^2} \int_\xph^1 \frac{\rd z}{z}
  \int^{\frac{\mu^2}{1-z}}_{\frac{m_p^2 x^2}{1-z}} \frac{\df Q^2}{Q^2} \frac{Q^4}{(Q^2+M_Z^2)^2} \bigg[ -z ^2\, F_L(x/z,Q^2) +\Bigl(
    z p_{\gamma q}(z) + \frac{2 x ^2 m_p^2}{Q^2} \Bigr)
    F_2(x/z,Q^2) \bigg]\, \nn
    & \quad +  \frac{g^2}{8 \pi^2} \int_\xph^1 \frac{\rd z}{z} z p_{\gamma q}(z) \left[ 
    \ln \frac{\mu^2 }{M_Z^2(1-z)+\mu^2} - \frac{M_Z^2(1-z)}{M_Z^2(1-z)+\mu^2}\right] F_2(x/z,\mu^2) \nn
    & \quad + \frac{g^2}{8 \pi^2}\int_\xph^1 \frac{\rd z}{z} (-z^2)
    F_2(x/z,\mu^2)
\,,\end{align}
%%%
which agrees with \eq{16.81}. The overall normalization differs by 8 because the coupling for the $W$ is $g/(2 \sqrt 2)$ rather than $g$.

%===============================================================================
\subsection{Longitudinal polarization}
%===============================================================================

\begin{figure}
\centering
\includegraphics[width=6cm]{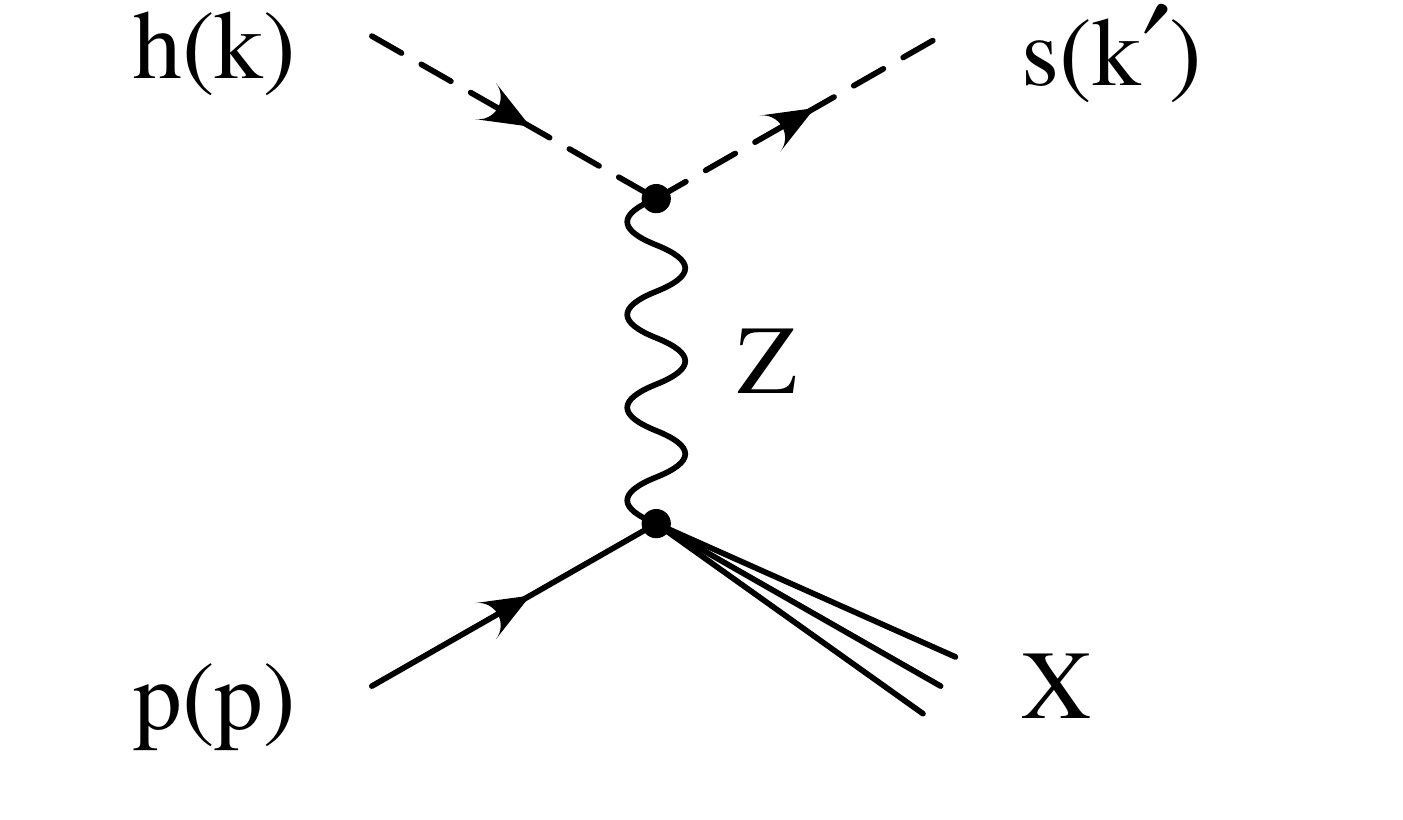}
\caption{{Feynman diagram for heavy scalar production via Higgs-proton scattering\ $h+p\rightarrow s+X$.}}
\label{fig:6}
\end{figure}

Longitudinal gauge boson PDFs present novel features, because they  only exist after spontaneous symmetry breaking. The first step is to identify a process to which they contribute at leading power, for which we consider 
%%%
\bea \label{eq:5.11a}
h(k) + p(p) \rightarrow s(k') + X 
\,,\eea
%%%
shown in fig.~\ref{fig:6}. Here the interaction between the $Z$ boson and scalars is described by the gauge invariant Lagrangian
%%%
\bea \label{eq:5.11}
\mathcal{L}_{\rm int} = \frac{\img}{\Lambda} \partial_\mu s \,\big[\Phi^\dagger D^\mu \Phi - (D^\mu \Phi^\dagger) \Phi \big],
\eea
%%%
where $\Phi$ is a charged scalar whose vacuum expectation value breaks the gauge symmetry, and $s$ is a heavy neutral scalar with mass $M_s$. After spontaneous symmetry breaking, 
%%%
\begin{align}
\Phi &= \frac{v + h}{\sqrt 2}
\end{align}
%%%
in unitary gauge, where $h$ denotes the Higgs field. Eq.~\eqref{eq:5.11} then becomes
%%%
\bea \label{eq:5.12}
\mathcal{L}_{\rm int} = -\frac{gv}{\Lambda}\, h\,\partial_\mu s \, Z^\mu.
\eea
%%%
Note that to obtain a term in which interactions with longitudinal gauge bosons are not power suppressed requires operators involving the Higgs field $\Phi$, resulting in interactions proportional to the vacuum expectation value $v$, as shown in \eq{5.12}.

The first step in obtaining the longitudinal $Z$ PDF is to factorize the cross section for the process in \eq{5.11a} in terms of structure functions, 
%%%
\bea
\sigma_{hp} &=&
 \frac{1}{4 \,p\cdot k} \int \frac{\df^4 q}{(2\pi)^4 }\,\frac{g^2}{(q^2-M_Z^2)^2}\,\big[4\pi \, W_{\mu\nu} S^{\mu\nu}\big]\, (2\pi)\,\delta[{(k-q)^2-M_s^2}] \, \theta(k^0-q^0) \nn
  &\quad& \times \theta[(p+q)^2- m_p^2]\ \theta(p^0+q^0)\,,
\eea
%%%
where the scalar tensor $S^{\mu \nu}$ is
%%%
\bea \label{eq:5.16}
S^{\mu\nu}(k) = \frac{g^2v^2}{\Lambda^2}  k'^\mu k'^\nu = \frac{g^2v^2}{\Lambda^2} (k-q)^\mu (k-q)^\nu \ .
\eea
%%%
The  scalar tensor couples only to longitudinally-polarized gauge bosons and not to transverse ones, so factorization directly gives  the longitudinal $Z$ PDF.
The scattering cross section in the limit $M_s \to \infty$ is given by
%%%
\begin{align}\label{eq:5.17}
\sigma_{hp} &=  \frac{g^4v^2}{16\pi \Lambda^2} \int_x^{1} \frac{\df z}{z} \int_{\frac{x^2 m_p^2}{1-z}}^{\frac{M_s^2 (1-z)}{z} } \frac{\df Q^2}{Q^2} \, \frac{Q^4}{(Q^2+M_Z^2)^2} \bigg\{\frac{z^2Q^2}{4M_s^4}\Big(1+\frac{M_s^2}{Q^2}\Big)^2F_L \left({x}/{z},Q^2\right) \nonumber\\
& \quad - \frac{1}{M_s^2}\bigg[z+\left(z-1\right)\frac{M_s^2}{Q^2} + \frac{x^2 m_p^2}{M_s^2}\Big(1+\frac{M_s^2}{Q^2}\Big)^2\bigg] F_2\left({x}/{z},Q^2\right)\biggr\} \ .
\end{align}
%%%
Splitting the $Q^2$ integral into two parts at $Q^2=\mu^2/(1-z)$, and neglecting power corrections gives
%%%
\begin{align}\label{eq:5.17b}
\sigma_{hp} &=  \frac{g^4 v^2 }{16 \pi \Lambda^2}\int_x^{1} \frac{\df z}{z}\bigg\{  \int_{\frac{x^2m_p^2}{1-z}}^{\frac{\mu^2}{1-z} }  \frac{ \df Q^2 }{(Q^2+M_Z^2)^2}\biggl[ \frac{z^2}{4} \, F_L \left({x}/{z},Q^2\right) + \Big(1-z-\frac{x^2m_p^2}{Q^2}\Big)\,F_2\left({x}/{z},Q^2\right)\biggr] \nn
&\quad  +\bigg[\frac{(1-z)^2}{M_Z^2(1-z)+{\mu^2}}\bigg]   F_2\left({x}/{z},\mu^2\right)\biggr\}\,.
\end{align}
%%%

Again, this cross section can also be written in terms of proton PDFs using \eq{5.8}.
In the limit $M_Z/M_s \rightarrow 0$, the $Z$ boson cross section $\hat \sigma_{hZ}$ is
%%%
\bea \label{eq:5.18}
\hat\sigma_{hZ}(x,\mu) &=&  \frac{\pi g^2}{4\Lambda^2}\frac{v^2}{M_Z^2} \delta(1-x) 
\eea
%%%
for longitudinally polarized $Z$ bosons, and power suppressed for transversely polarized $Z$ bosons. Thus \eq{5.8} picks out the longitudinal $Z$ PDF.
The contribution from quarks is calculated from tree-level Higgs-quark scattering via $Z$ exchange, and is 
%%%
\bea
\hat\sigma_{hq}(z,\mu)  = \frac{g^4 v^2 z}{16 \pi \Lambda^2 M_s^2}\left[\frac{2}{\epsilon}  - \ln\frac{M_s^2(1-z)^2}{z\, M_Z^2}\right] \,,
\eea 
%%%
where $z=M_s^2/ \hat s$, and is power suppressed relative to \eq{5.18}.  From \eq{5.17b} and \eq{5.18}, we get the longitudinal $Z$  boson PDF 
%%%
\begin{align}
x f_{Z_L}(x,\mu) &= \frac{g^2 M_Z^2}{4 \pi^2}\int_x^{1} \frac{\df z}{z} \int_{\frac{x^2m_p^2}{1-z}}^{\frac{\mu^2}{1-z} }  \frac{ \df Q^2 }{(Q^2+M_Z^2)^2} \biggl[ \Big(1-z-\frac{x^2m_p^2}{Q^2}\Big)\,F_2\left({x}/{z},Q^2\right) + \frac{z^2}{4} \, F_L \left({x}/{z},Q^2\right)\biggr] \nn
& \quad +  \frac{g^2 M_Z^2}{4\pi^2}\int_x^{1} \frac{\df z}{z}\ F_2\left({x}/{z},\mu^2\right)  \biggl[\frac{(1-z)^2}{M_Z^2(1-z)+\mu^2}\biggr] \,.
\end{align}
%%%
This agrees with \eq{47} taking into account the overall factor of $1/8$, as in \eq{5.10}.

To simplify the presentation, the calculations in this section have been done for a spontaneously broken $U(1)$ gauge theory. However, it should be clear how these can be extended to the case of a spontaneously broken $SU(2) \times U(1)$ in the Standard Model.

%%%%%%%%%%%%%%%%%%%%%%%%%%%%%%%%%%%%%%%%%%%%%%%%%%%%%%%%%%%%%%%%%%%%%%%%%%%%%%%%
\section{Numerics}\label{sec:numerics}
%%%%%%%%%%%%%%%%%%%%%%%%%%%%%%%%%%%%%%%%%%%%%%%%%%%%%%%%%%%%%%%%%%%%%%%%%%%%%%%%

%%% FIG %%%
\begin{figure}
\centering
\includegraphics[width=7.5cm]{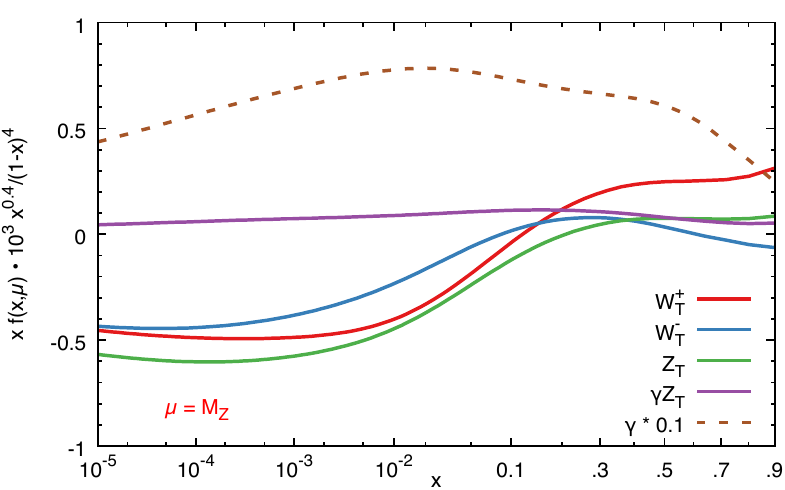}\hspace{0.25cm}
\includegraphics[width=7.5cm]{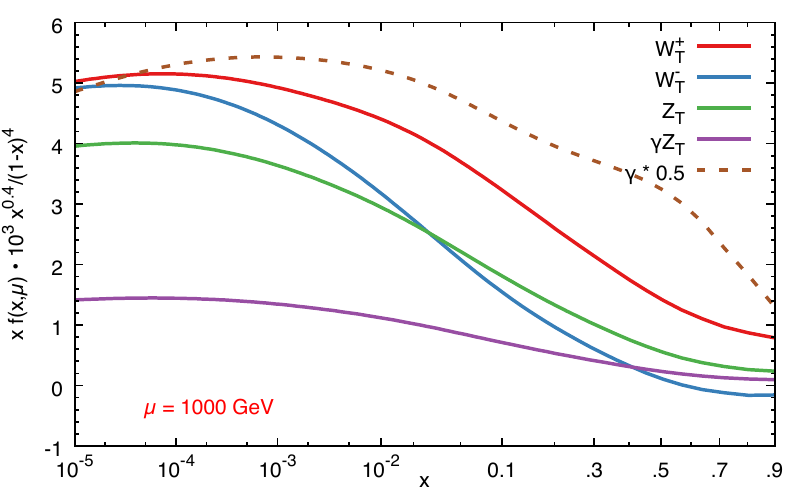}
\caption{\label{fig:num1} The transverse gauge boson PDFs $f_{W^+_T}$ (red), $f_{W^-_T}$ (blue), $f_{Z_T}$  (green) and $f_{\gamma Z_T}$ (purple) for $\mu=M_Z$ and $\mu=1000$\,GeV. The unpolarized photon PDF (dashed, brown) has also been shown for comparison, multipled by $0.1$ at $\mu=M_Z$ and by $0.5$ at $\mu=1000$\,GeV, so it fits on the same plot.}
\end{figure}
%%%
%
%%% FIG %%%
\begin{figure}
\centering
\includegraphics[width=7.5cm]{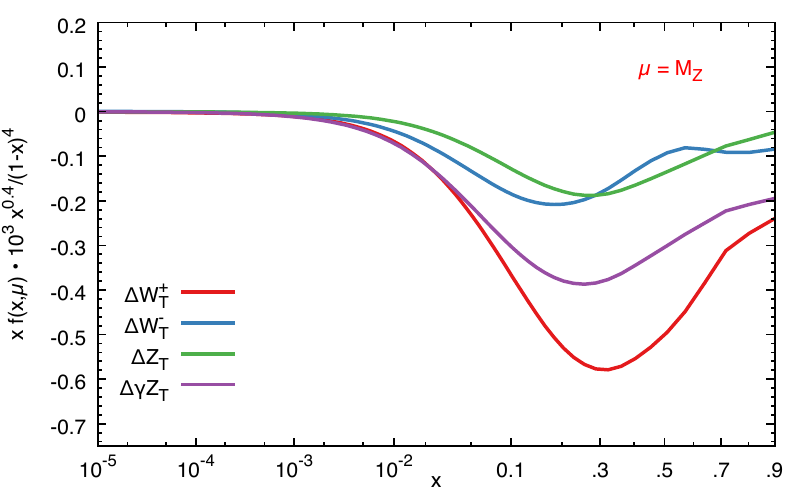}\hspace{0.25cm}
\includegraphics[width=7.5cm]{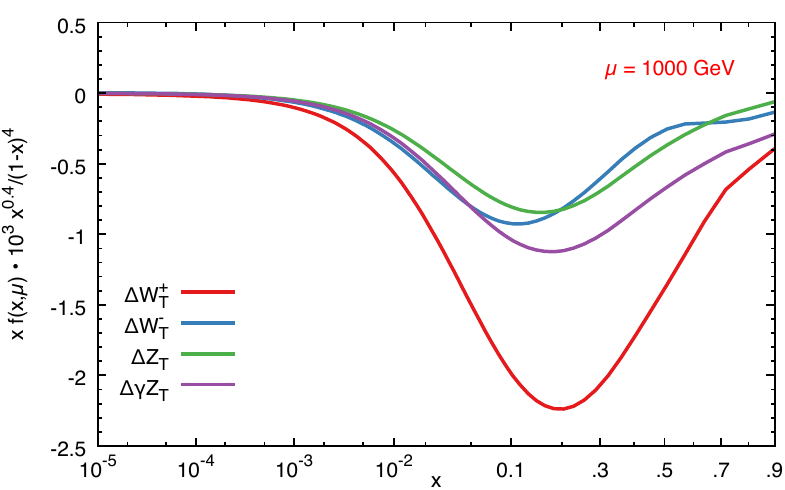}
\caption{\label{fig:num2} The polarized gauge boson PDFs $f_{\Delta W^+_T}$ (red), $f_{\Delta W^-_T}$ (blue), $f_{\Delta Z_T}$  (green) and $f_{\Delta \gamma Z_T}$ (purple) for $\mu=M_Z$ and $\mu=1000$\,GeV. }
\end{figure}
%%%
%
%%% FIG %%%
\begin{figure}
\centering
\includegraphics[width=7.5cm]{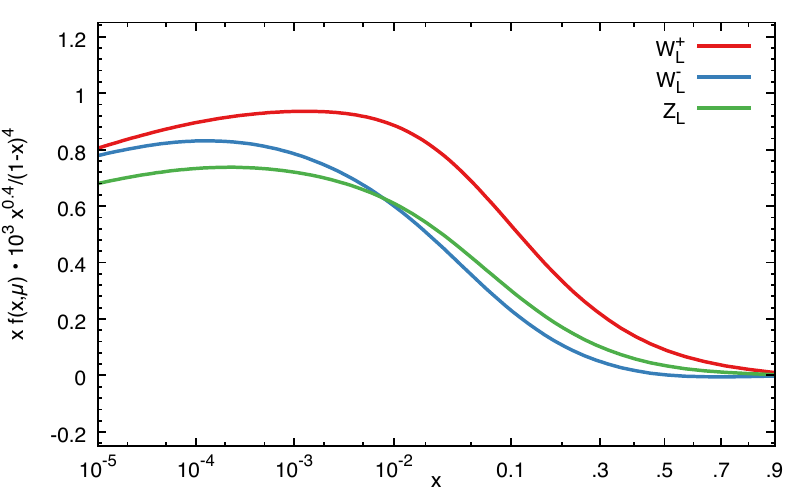}
\caption{\label{fig:num3} The longitudinal gauge boson PDFs $f_{W^+_L}$ (red), $f_{W^-_L}$ (blue), and $f_{Z_L}$  (green) for $\mu=M_Z$. The longitudinal PDF does not depend on $\mu$ to the order computed in the plot.}
\end{figure}

In this section we present numerical results for the electroweak gauge boson PDFs, obtained using eqs.~\eqref{eq:16.81}, \eqref{eq:16.9}, \eqref{eq:16.8}, \eqref{eq:16.10}, and \eqref{eq:47}. The equations have corrections of order $\alpha^2_2$, arising from e.g.~the graphs in \fig{2}. All QCD corrections and $m_p^2/M_W^2$ power corrections are included automatically by using the deep-inelastic scattering structure functions. 

The expressions for the electroweak gauge boson PDFs involve integrations over $Q^2$ between $m_p^2 x^2/(1-z)$ and $\mu^2/(1-z)$, and thus include the elastic scattering and resonance regions. Compared to the photon PDF, the integrands have factors of $Q^2/(Q^2+M^2)$, where $M=M_W,M_Z$. Thus the low-$Q^2$ part of the integration region is suppressed by  $\sim m_p^2/M_W^2 \sim 10^{-4} $, the size of low-energy weak interaction corrections and smaller than the missing $\alpha^2_2$ corrections, so we only need values of the structure functions for $Q^2$ of order the electroweak scale. The $x$ integral still includes elastic scattering at $x/z=1$, but for large $Q^2$ the elastic  form-factors are power suppressed. This justifies using the expressions for the $F_2$ and $F_3$ structure functions at lowest order QCD in terms of PDFs in \eq{2.17}, \eq{2.17c}, and setting $F_L$ to zero (since it starts at order $\alpha_s$), to evaluate the PDFs. This method is not as accurate as using the experimentally measured structure functions, because it introduces order $\alpha_s(M_W)$ radiative corrections as well as $m_p^2/M_W^2$ power corrections.\footnote{The radiative corrections depend on $\alpha_s$ and $\alpha_2$ evaluated at $Q^2$ scales that contribute to the integral. One can minimize these corrections by including higher-order terms in the expressions for $F_i$. If, instead, the experimentally measured structure functions are used, the corrections depend on $\alpha_2(\mu)$, where $\mu > M_Z$ is the scale at which the PDF is evaluated.}

The numerical integrations are done using the PDF set {\tt NNPDF31\_nlo\_as\_0118\_luxqed} PDFs~\cite{Bertone:2017bme} and the
 LHAPDF~\cite{Buckley:2014ana} and {\tt ManeParse}~\cite{Clark:2016jgm} interfaces.  A detailed numerical analysis including PDF evolution and errors is beyond the scope of this paper. The results for electroweak gauge boson PDFs are shown in \fig{num1}, \ref{fig:num2}, and \ref{fig:num3} for $\mu=M_Z$ and $\mu=1000$\,GeV using the {\tt NNPDF31} central PDF set. The electroweak PDFs have been renormalized  in the \MSbar\ scheme, so they do not have to be positive. They start at order $\alpha_2$, rather than order unity, and NLO corrections can be negative.
 
The transverse PDFs (\fig{num1}) are small at $\mu=M_Z$, and rapidly grow with energy to be almost comparable to the photon at $\mu=1000$\,GeV due to the evolution in \eq{2.34}.  The $W^+_T$ PDF is larger than $W^-_T$, since $f_u > f_d$ in the proton.  The PDFs are negative at $\mu=M_Z$, but rapidly become positive as the $\ln Q^2$ part dominates over the \MSbar\ subtraction term. The polarized PDFs (\fig{num2}) are negative, since quark PDFs are larger than antiquark PDFs, and left-handed quarks prefer to emit left-hand circularly polarized $W$ bosons. The longitudinal PDFs (\fig{num3}) are comparable to the transverse ones at $\mu=M_Z$. Since the longitudinal PDFs are $\mu$-independent, only one plot has been shown. The transverse PDFs rapidly become larger than the longitudinal ones as $\mu$ increases.

\section{Conclusions}

We have computed the electroweak gauge boson PDFs at a scale $\mu \sim M_{W,Z} \gg m_p$ for transversely and longitudinally polarized gauge bosons, by computing the proton matrix element of the PDF operator in terms of proton structure functions for charged and neutral current scattering.
The PDFs can be evolved to higher energies using the evolution equations derived recently in refs.~\cite{Bauer:2017isx,Manohar:2018kfx}. The electroweak gauge boson PDFs have been computed previously using the effective $W$ approximation~\cite{Chanowitz:1984ne,Dawson:1984gx,Kane:1984bb}. The leading logarithmic piece of our result agrees with their expressions, but the full order $\alpha$ results differ. Numerical values for the PDFs at the representative scales $\mu=M_Z$ and $\mu=1000$\,GeV are given in sec.~\ref{sec:numerics}. More detailed numerical results are beyond the scope of this paper, and will be given elsewhere.

\begin{acknowledgments}
This work is supported by the DOE grant DE-SC0009919, the ERC grant ERC-STG-2015-677323,  the D-ITP consortium, a program of the Netherlands Organization for Scientific Research (NWO) that is funded by the Dutch Ministry of Education, Culture and Science (OCW),  and the Munich Institute for Astro- and Particle Physics (MIAPP) of the DFG cluster of excellence ``Origin and Structure of the Universe."
\end{acknowledgments}

\phantomsection
\addcontentsline{toc}{section}{References}
\bibliographystyle{jhep}
\bibliography{wpdf}

\end{document}